\documentclass[10pt,a4paper]{article} 

\usepackage{geometry}
\usepackage{times} 
\usepackage{graphicx}
\usepackage{amsmath}
\usepackage{natbib}
\usepackage{threeparttable}
\usepackage{multirow}
\usepackage{subfigure}

\newcommand{\eg}{e.g.,}
\newcommand{\Eg}{E.g.,}
\newcommand{\etal}{et al.}
\newcommand{\ie}{i.e.,}

\newcommand{\sect}[1]{Section \ref{s:#1}}
\newcommand{\eqn}[1]{Eq.\ (\ref{e:#1})}
\newcommand{\Eqn}[1]{Equation (\ref{e:#1})}
\newcommand{\fig}[1]{Fig.\ \ref{f:#1}}
\newcommand{\Fig}[1]{Figure \ref{f:#1}}
\newcommand{\tbl}[1]{Table \ref{t:#1}}
\newcommand{\code}[1]{\texttt{#1}}

\newcommand{\hide}[1]{} 

\def\paper#1 #2 #3 #4 #5 #6 {#1, #2. #3. #4\ #5, #6.}
\def\inpress#1 #2 #3 #4 {#1, #2. #3. #4, in press.}
\def\preprint#1 #2 #3 #4 {#1, #2. #3. ArXiv e-prints #4.}
\def\submitted#1 #2 #3 #4 {#1, #2. #3. #4, submitted.}
\def\inprep#1 #2 #3 #4 {#1, #2. #3. #4, in preparation.}
\def\thesis#1 #2 #3 #4 #5 {#1, #2. #3. Thesis, #4. #5 pp.}
\def\book#1 #2 #3 #4 {#1, #2. #3. #4.}
\def\chap#1 #2 #3 #4 #5 #6 #7 {#1, #2. #3. In: #4 (Eds.), #5. #6, pp.\ #7.}
\def\aIII#1 #2 #3 {#1, 2002. #2. In: Bottke Jr., W.F. \etal\ (Eds.), Asteroids III. Univ.\ of Arizona
  Press, Tucson, pp.\ #3.}

\begin{document}

\begin{center}
\textbf{\Large Structural analysis of rubble-pile asteroids applied to collisional evolution}\\
\bigskip
\bigskip

Yang Yu\footnote{Corresponding Author, School of Aeronautic Sci. \& Engr., Beihang Univ., 100191 Beijing, CHN, Email: \code{yuyang.thu@gmail.com}.}, 
Derek C. Richardson$^2$, 
Patrick Michel$^3$\\ 
\bigskip
$^1$Beihang University, 100191 Beijing, China\\
$^2$Department of Astronomy, University of Maryland, College Park MD 20740-2421 United States\\
$^3$Observatoire de la C\^ote d'Azur, UMR 7293 Lagrange/CNRS, 06304 Nice Cedex 4, France\\
\bigskip
\bigskip
\end{center}

\noindent
\textbf{Abstract:} Solar system small bodies come in a wide variety of shapes and sizes, which are achieved following very individual evolutional paths through billions of years. Nevertheless, some common mechanisms can still be found during these processes, \eg\ rubble-pile asteroids tend towards fluid equilibrium as they are reshaped by external disturbances. This paper focuses on the reshaping process of rubble-pile asteroids driven by meteorite impacts. A mesoscale cluster of solid spheres is employed as the principal model for a rubble-pile asteroid, for which little is actually known about their interior structure. We take this simple model as a rough guide to the qualitative aspects of the reshaping processes, and it can reveal, to some degree, the inner workings of rubble-pile asteroids. In our study, numerous possible equilibrium configurations are obtained via Monte Carlo simulation, and the structural stability of these configurations is determined via eigen analysis of the geometric constructions. The eigen decomposition reveals a connection between the cluster's reactions and the types of external disturbance. Numerical simulations are performed to verify the analytical results. The gravitational $N$-body code \code{pkdgrav} is used to mimic the responses of the cluster under intermittent non-dispersive impacts. We statistically confirm that the stability index $I_\textup{S}$, the total gravitational potential $P_\textup{G}$ and the volume of inertia ellipsoid $V_\textup{E}$ show consistent tendency of variation. A common regime is found in which the clusters tend towards crystallization under intermittent impacts, \ie\ only the configurations with high structural stability survive under the external disturbances.The results suggest the trivial non-disruptive impacts might play an important role in the rearrangement of the constituent blocks, which may strengthen these rubble piles and help to build a robust structure under impacts of similar magnitude. The final part of this study consists of systematic simulations over two parameters, the projectile momentum and the rotational speed of the cluster. The results show a critical value exists for the projectile momentum, as predicted by theory, below which all clusters become responseless to external disturbances; and the rotation proves to be significant for it exhibits an ``enhancing" effect on loose-packed clusters, which coincides with the observation that several fast-spinning asteroids have low bulk densities.

\noindent
\textbf{Keywords:} Asteroids, interior structure; geological processes; N-body simulations; gravitational aggregates; meteorite impacts.

\bigskip
\bigskip

\section{Introduction} \label{s:intro}

There is abundant evidence that most asteroids between $\sim100$ m and $\sim100$ km in size are gravitational aggregates of boulders, whether born as loosely consolidated structures or formed from the remains of shattered larger bodies [1]. The constitution of asteroids proved to have a significant role during their subsequent evolution [2,3]. Generally, rubble-pile asteroids are very responsive to close planetary tides because of the near-strengthless interiors [4,5], and present good resistance to further disruption from meteorite impacts due to the high energy diffusion of their structures [6]. Observations have shown reshaping marks of rubble-pile asteroids due to the disturbances during geological processes. More than $20$ asteroid families in the main belt suggest a collisional origin in which large impactors shattered the parent body and the fragments reaccumulated to form the large family members [7]. Comet breakups, \eg Comet D/Shoemaker-Levy 9 at Jupiter, showed the tidal disruption effects of planetary tides during close encounter [8,9]. Richardson \etal\ used $N$-body simulations to explore the possible outcomes of tidal encounters of rubble-pile asteroids with the Earth, and found similar fragment trains are created for severe disruptions [10].

In reality, the violent processes have statistically low probability, while moderate disturbances, such as weak tidal encounters or non-dispersive impacts from micrometeorites, seem to be more common experiences for most rubble-pile asteroids. These events can lead to seismic shaking and slight distortion of the interior structure. It was found that the double-lobed shape of several asteroids such as (4769) Castalia, (2063) Bacchus, (4179) Toutatis and (216) Kleopatra could be formed by the gentle mutual reactions between similar-sized bodies following a tidal breakup or a low-speed collision [10--12]. Accordingly, the continual low-energy events may have the potential to modify the interior structure of rubble-pile asteroids in quasi-equilibria. Early work suggested that asteroids may have fluid equilibrium shapes [13], but more recent observations show that this is generally not the case [14].  In fact, asteroids $\sim100$ km or smaller are likely formed of discrete blocks with different sizes and shapes, which provides some shear strength that prevents the overall shape from coming to fluid equilibrium [15,16]. Tanga \etal\ proposed that a general mechanism exists to produce the observed shape distribution, that the fragmented structure could progressively reshape via gradual rearrangement of the constituent blocks due to external forces [17], such as planetary tides, non-dispersive impacts etc. $N$-body simulations were employed to investigate the equilibration of rubble piles based on a $4$-dimensional parametric space, and the results showed external factors can break the interlocking of the rubble piles and drive them to asymptotically evolve toward fluid equilibrium. 

It is interesting to develop this idea further, to explore how the reshaping process of a rubble pile depends on its interior structure, \ie\ can we figure out the structural stability and collisional responses of a rubble pile from its configuration? One approach to start with a basic model. We choose a simple soft-sphere cluster to mimic the real rubble pile, with all blocks modeled as solid spheres [18,19]. This is not for accurate reproduction of the asteroid constitution but serves as a bridge to understanding the natural evolution of rubble-pile asteroids during intermittent disturbances. Richardson \etal\ proposed a scheme to place the asteroid in a two-parameter diagram of porosity and relative tensile strength, defining eight typical asteroid structures: monolithic, fractured, shattered, shattered with rotated components, rubble pile, coherent rubble pile, weak-and-porous and strong-and-porous [1]. One benefit of this classification is it helps distinguish the responses of different types of interior structures during geological processes. In this study, we focus on clusters of tens of spheres based on the following rationale: first, the mesoscale clusters can be regarded as representative of some (perhaps many) asteroids that have not been fully shattered, and their overall shapes are largely determined by some big components; second, the rubble piles composed of big constituents  behave more like discrete bodies than fluid bodies and provide a good approximation to real rubble-pile asteroids; third, rubble piles of tens of spheres may have been representative for the complex dynamics (though not as complex as the aggregates of more spheres) and possess various possible equilibrium configurations. This diversity of this simple case essentially relies on the nature of sphere clusters, thus it holds even when some complex interactions, such as the geometrical locking and static friction, are omitted. 

This paper presents a quantitative estimation of the structural stability of the rubble-pile model, which is then applied to the investigation of the responses of rubble-pile asteroids to different disturbances. Numerical simulations are performed using the code \code{pkdgrav}, a gravitational $N$-body code which is well suited for discrete material. \sect{eigen} introduces the theoretical analysis of structural stability of a rubble-pile configuration, along with a numerical example as validation. \sect{experi} describes the three groups of simulations we performed, including choice of parameters used. \sect{result} discusses the results of these numerical experiments. We validate the results of our structural analysis in two ways: to check the responses of numerous sample clusters statistically, and to track the detailed behaviors of a specified cluster.

\section{Structural Analysis of Equilibrium Cluster} \label{s:eigen}

\subsection{Representation of Equilibrium Cluster} \label{s:repres}

The mesoscale clusters are obtained through an assembly of equal-size soft spheres. No cohesion or friction is considered here, \ie\ the spheres stay together via mutual gravity alone. Initialized as a random cloud of particles in space, the spheres evolve gradually into an equilibrium cluster due to the frictional damping and collisional dissipation. Consider a cluster of $N$ spheres; the mass and radius of each sphere are defined as $m$ and $R$, and $\mathbf{r}_i$ is the position vector of sphere $i$. Define the delta function

\begin{equation}
\label{e:delta}
\delta_{i,j} = \left\{\begin{matrix}
1 & \mathrm{if} \ \left| \mathbf{r}_i - \mathbf{r}_j \right| < 2R\\
0 & \mathrm{if} \ \left| \mathbf{r}_i - \mathbf{r}_j \right| \geq 2R\\
\end{matrix}\right. 
\end{equation}

\noindent and represent the position of sphere $\mathbf{r}_i$ in the central principal frame (CPF), which is defined with the origin at the system mass center, and $x$-, $y$-, $z$-axes aligned along the small, medium and large principal axes of inertia, respectively. Note that CPF exists only if the cluster is equilibrated. The equilibrium equation of sphere $i$ is given by \eqn{equeqn}. 

\begin{equation}
\label{e:equeqn}
\sum_{j \neq i} \mathbf{G}_{i,j} + \sum_{j \neq i, \delta_{i,j} = 1} \mathbf{C}_{i,j}  + \Omega^2 \mathbf{r}_i - (\mathbf{\Omega} \cdot \mathbf{r}_i) \mathbf{\Omega} = \mathbf{0}, \ \ i = 1, 2, \cdots, N, 
\end{equation}

\begin{equation}
\label{e:grveqn}
\mathbf{G}_{i,j} = \frac{\textup{G} m^2}{\left| \mathbf{r}_i - \mathbf{r}_j \right|^3} \left( \mathbf{r}_j - \mathbf{r}_i \right), 
\end{equation}

\begin{equation}
\label{e:cfeqn}
\mathbf{C}_{i,j} = -\textup{K}_n \left( 1 - \frac{2 R}{\left| \mathbf{r}_i - \mathbf{r}_j \right|} \right) \left( \mathbf{r}_i - \mathbf{r}_j \right) + \textup{K}_t \mathbf{S}_{i,j}.
\end{equation}

\Eqn{grveqn} and \Eqn{cfeqn} define respectively the gravitational attraction and contact force on sphere $i$ from sphere $j$. And $\mathbf{S}_{i,j}$ indicates the tangential displacement between sphere $i$ and $j$ (see Ref. [19] for a detailed description). The angular velocity $\mathbf{\Omega}$ is a constant vector if the cluster is equilibrated. $\textup{G}$ indicates the gravitational constant, and $\textup{K}_n$, $\textup{K}_t$ indicate the constants for the normal spring and tangential spring, respectively. Then every solution $\mathbf{x}$ to this 3$N$-dimensional equation presents a possible equilibrium configuration of this cluster, and is called the configuration for short, \ie\

\begin{equation}
\label{e:config}
\mathbf{x} = \left [ \mathbf{r}_1, \mathbf{r}_2, \cdots, \mathbf{r}_N \right ] ^ T, 
\end{equation}

and as a canonical expression of the cluster in CPF, $\mathbf{x}$ represents a unique structure of the cluster, independent of the translation/rotation with respect to the inertia frame. Denoting the left-hand side of \eqn{equeqn} as $\mathbf{F}_i$, with the three components $F_{i,x}, F_{i,y}, F_{i,z}$, \eqn{equeqn} can be written as

\begin{equation}
\label{e:equeqn2}
F_{i,x} = 0, F_{i,y} = 0, F_{i,z} = 0, \ \ i = 1, 2, \cdots, N.
\end{equation}

The Jacobi matrix of \eqn{equeqn2} yields 

\begin{equation} 
\label{e:LinMatx}
\mathbf{M} = \left [ \frac{\partial F_{1,x}}{\partial \mathbf{x}}, \frac{\partial F_{1,y}}{\partial \mathbf{x}}, \frac{\partial F_{1,z}}{\partial \mathbf{x}}, \frac{\partial F_{2,x}}{\partial \mathbf{x}}, \frac{\partial F_{2,y}}{\partial \mathbf{x}}, \frac{\partial F_{2,z}}{\partial \mathbf{x}}, \cdots, \frac{\partial F_{N,x}}{\partial \mathbf{x}}, \frac{\partial F_{N,y}}{\partial \mathbf{x}}, \frac{\partial F_{N,z}}{\partial \mathbf{x}} \right ]. 
\end{equation}

The $3N \times 3N$ matrix $\mathbf{M}$ expresses the linear part of \eqn{equeqn} and determines the local behavior of this dynamical system, \ie\ the responses of the cluster after being disturbed from equilibrium configuration $\mathbf{x}$, as shown by the linear system of \eqn{respon}:

\begin{equation} 
\label{e:respon}
\begin{bmatrix} \mathbf{\dot y} \\
		            \mathbf{\ddot y}
\end{bmatrix}      =      \begin{bmatrix} \mathbf{O} & \mathbf{I}\\
   									   \mathbf{M} & \mathbf{O}
					   \end{bmatrix} \begin{bmatrix} \mathbf{y} \\
												     \mathbf{\dot y} 
									\end{bmatrix}, 
\end{equation}

\noindent where $\mathbf{y} = \delta \mathbf{x}$ is the disturbance around configuration $\mathbf{x}$, $\mathbf{O}$ is a $3N \times 3N$ zero matrix, and $\mathbf{I}$ is a $3N \times 3N$ identity matrix. The linear matrix of \eqn{respon} (symbolized as $\mathbf{L}$) presents the transient response of the cluster under any type of disturbance, either on the position or on the velocity, which enables us to check the structural stability of given configuration $\mathbf{x}$ via eigen analysis of $\mathbf{L}$ [20]. In the remaining part of this section, we will detail this correlation with an example. 

\subsection{An Example} \label{s:examp}

A mesoscale cluster of $27$ spheres is used as an example, which is strength-dominated in structure. The radius of the sphere is $30.0$ m. The normal chondrite density $\sim2.7$ g/cc is adopted here and the overall dimension is $\sim300$ m. To generate a configuration, spheres were randomly dispersed in a cubic region initially, and then they gathered under mutual gravity naturally and collapse to form an aggregate. Note that static friction in the equations of motion is omitted so that \eqn{respon} describes a continuous differentiable system, \ie\ the theory of linearized stability applies [21]. Further setting the tangential damping constant $\textup{C}_t$ and stiffness constant $\textup{K}_t$ both to zeros, the lack of friction makes the soft-sphere system more fluid, so the possible resulting configurations should be concentrated around the fluid equilibrium. The rotational state of each single sphere is neglected because the librational motion is decoupled from the translational motion for spheres. $162$ eigenvalues $\lambda_i \left ( i=1, \cdots, 162 \right )$ of the linearized matrix $\mathbf{L}$ are calculated at a given equilibrium configuration $\mathbf{x}$. 

\begin{figure}[h]
\centering
 \subfigure[Configuration]{
   \label{}
   \includegraphics[width=0.22\textwidth] {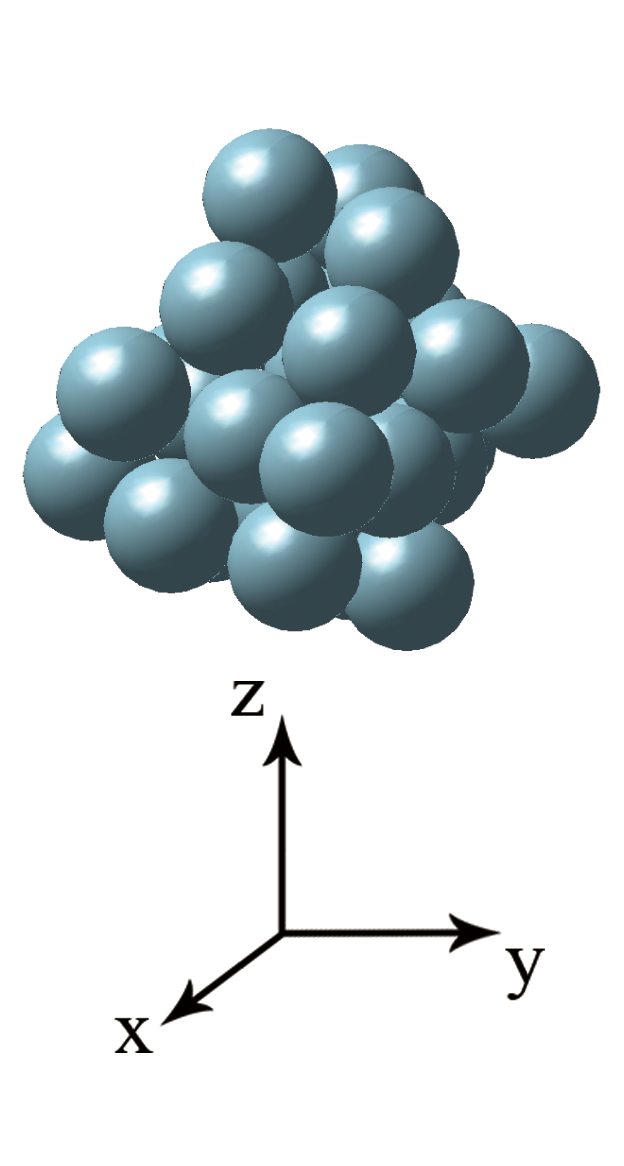}
 }
 \subfigure[The distribution of eigenvalues]{
   \label{}
   \includegraphics[width=0.52\textwidth] {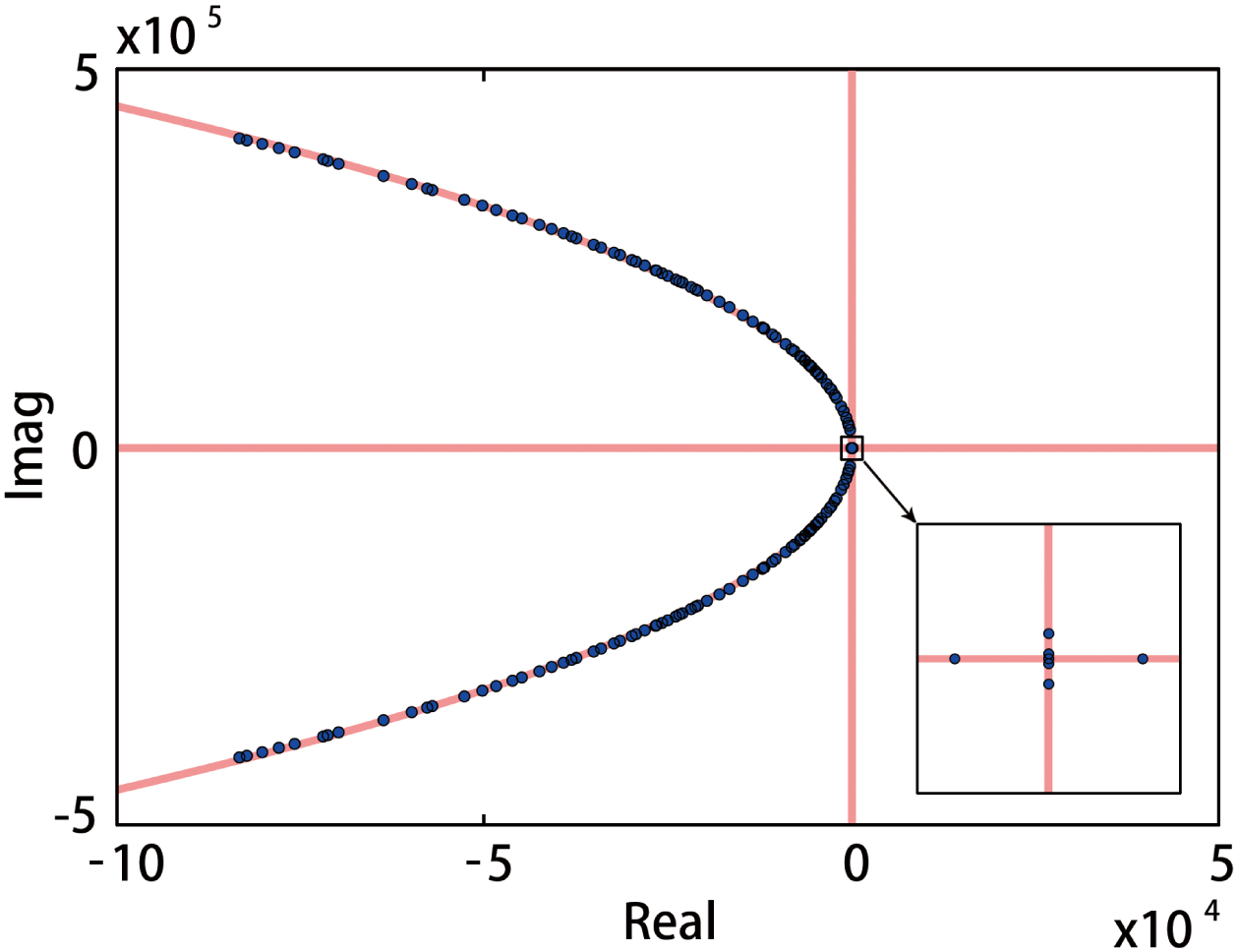} 
 }
\caption{The distribution of eigenvalues $\lambda_i$ on the complex plane (unit: s$^{-1}$) for given equilibrium configuration. (a) shows the configuration of given cluster, and (b) shows $\lambda_i$ on the complex plane. The light red lines indicate three feasible branches, and the solid circles indicate coordinates of $\lambda_i$. The floating box in (b) shows a enlarged view of the neighborhood around the origin. }
\label{f:eigenvalue}
\end{figure}

\Fig{eigenvalue} illustrates the distribution these eigenvalues (b) for given configuration (a). The $162$ eigenvalues are distributed in three branches on the complex plane: the stable branch is composed of the negative real axis and the parabola symmetric about it; the critical branch is the imaginary axis; and the unstable branch is the positive real axis. 

We confirm numerically the pattern of this distribution is an common characteristic of equilibrium configurations for any clusters (see also \sect{correl}). And it is correlated with the mechanical nature of the soft sphere element. To understand this, several points should be stated first: 

\begin{enumerate}

\item {Eigenvalues $\lambda_i$ are in pairs, \eg\ conjugate complex numbers (stable branch), conjugate imaginary numbers (critical branch) and opposite real numbers (positive value on unstable branch, negative value on stable branch). In physics, conjugate complex $\lambda_i$ with negative real part indicates the disturbed cluster experiences damped shaking to revert to the original configuration; conjugate imaginary $\lambda_i$ indicates constant-amplitude shaking; negative real $\lambda_i$ indicates directly reverting to the original configuration; and positive $\lambda_i$ indicates leaving the original configuration, \ie\ the disturbance is irreversible for given cluster.} 

\item {The reason for the parabolic stable branch of $\lambda_i$ is the linear mechanical properties in the soft-sphere element, \ie\ the constant damping coefficient $\textup C_n$ and the stiffness coefficient $\textup K_n$. For an arbitrary sphere of the cluster, the damping force and elastic force acting on it accumulate with the number of neighboring spheres, thus the ratio keeps constant and equals $\textup C_n/ \textup K_n$. Also we notice the real part of eigenvalue $\textup{Re} \left ( \lambda_i \right ) \propto  \textup C_n$ and the imaginary part $\textup{Im} \left ( \lambda_i \right ) \propto \sqrt{\textup K_n}$ [22], therefore $\textup{Re} \left ( \lambda_i \right )$ and $\textup{Im} \left ( \lambda_i \right )^2$ always change in proportion.} 

\item {Six zero eigenvalues always exist for the cluster, which corresponds to the six degrees of freedom for the rigid motion, \ie\ coherent translation or rotation of the given equilibrium configuration. Evidently, these types of disturbances lead to no response of the cluster configuration.} 

\end{enumerate} 

\begin{figure}[h]
\centering
 \subfigure[]{
   \label{}
   \includegraphics[width=0.22\textwidth] {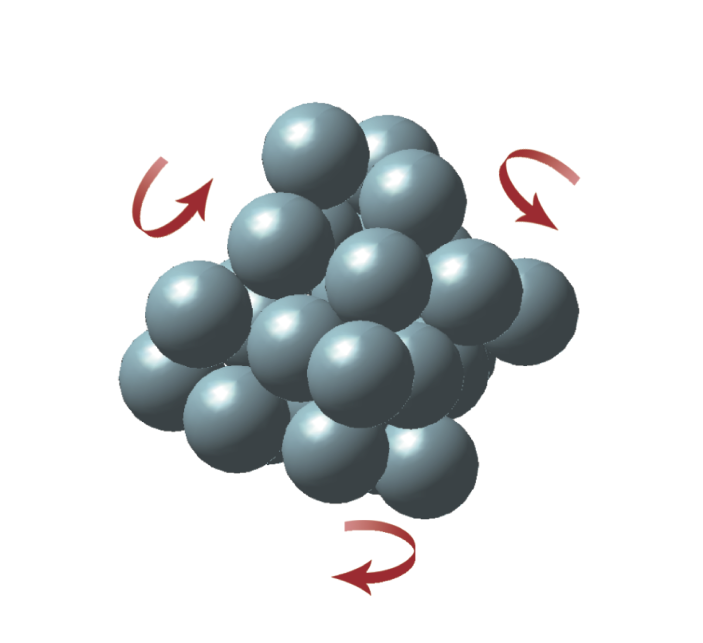}
 }
 \subfigure[]{
   \label{}
   \includegraphics[width=0.22\textwidth] {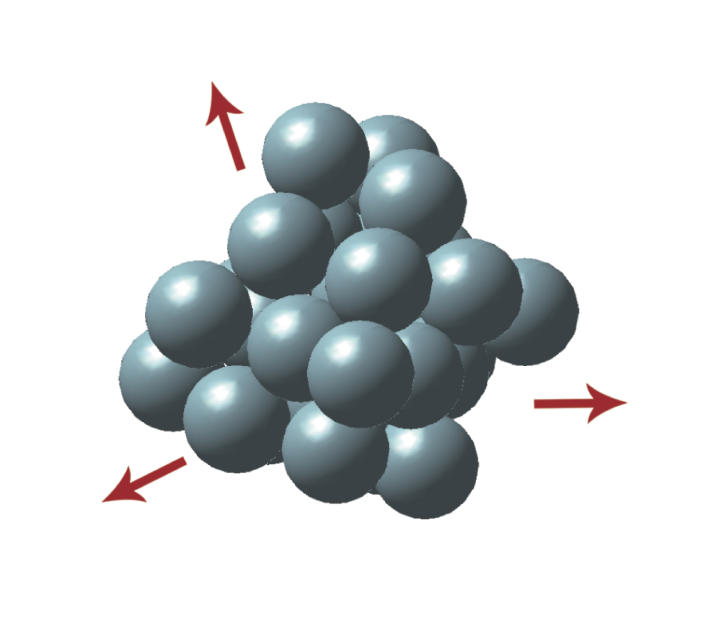} 
 }
 \subfigure[]{
   \label{}
   \includegraphics[width=0.22\textwidth] {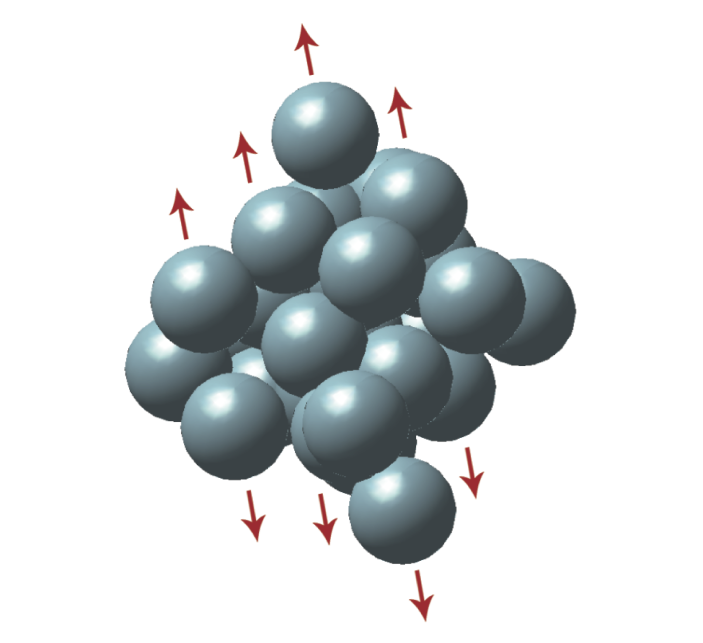} 
 }
 \subfigure[]{
   \label{}
   \includegraphics[width=0.22\textwidth] {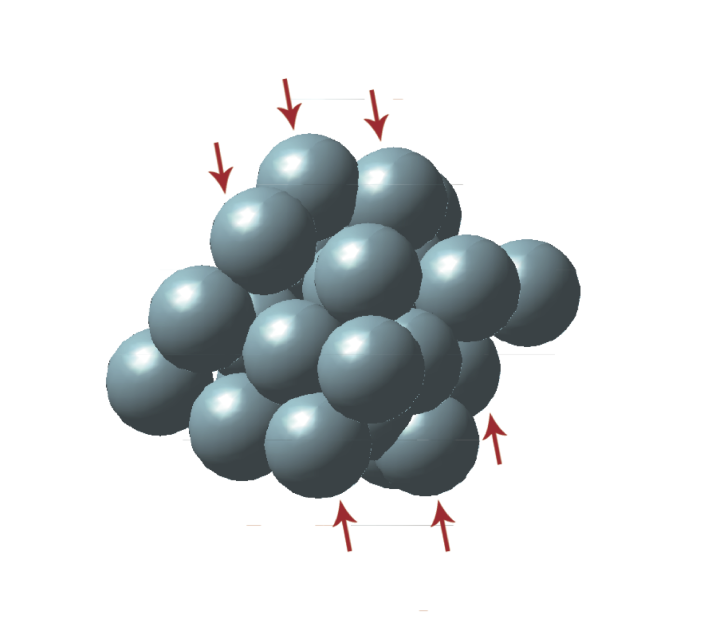}
 }
 \subfigure[]{
   \label{}
   \includegraphics[width=0.22\textwidth] {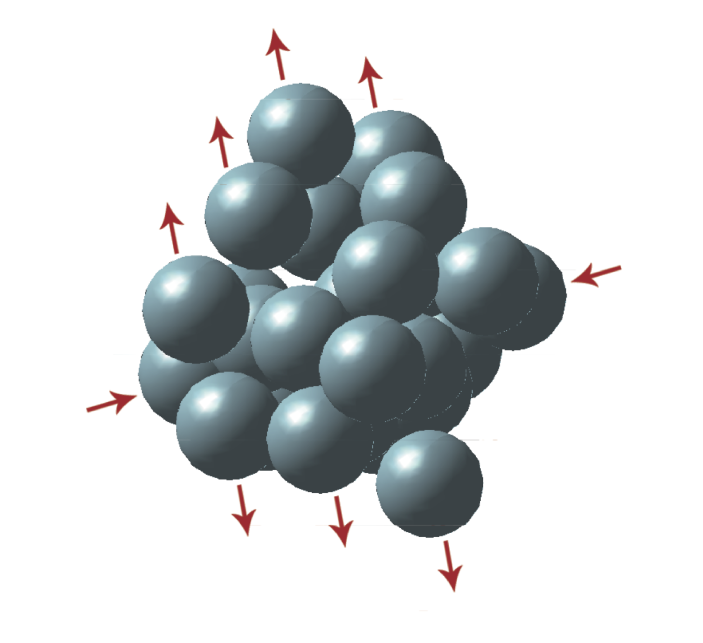} 
 }
 \subfigure[]{
   \label{}
   \includegraphics[width=0.22\textwidth] {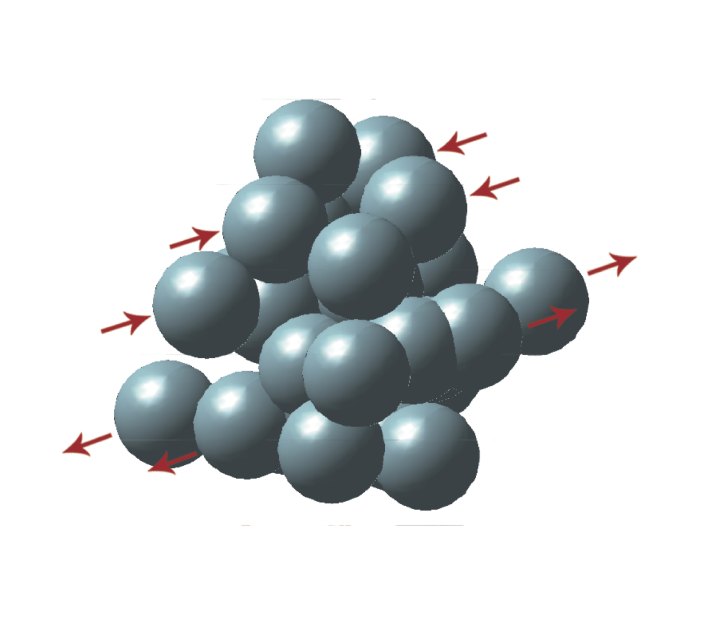}
 }  
 \subfigure[]{
   \label{}
   \includegraphics[width=0.22\textwidth] {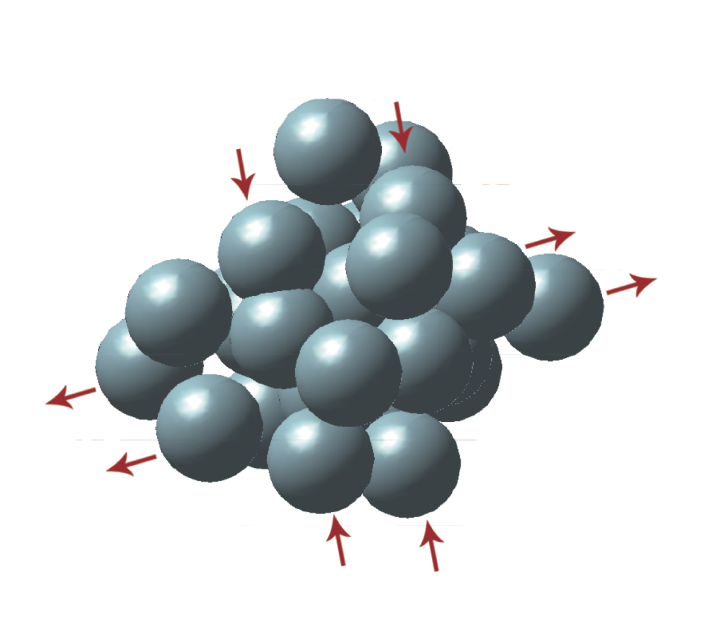}
 }
 \subfigure[]{
   \label{}
   \includegraphics[width=0.22\textwidth] {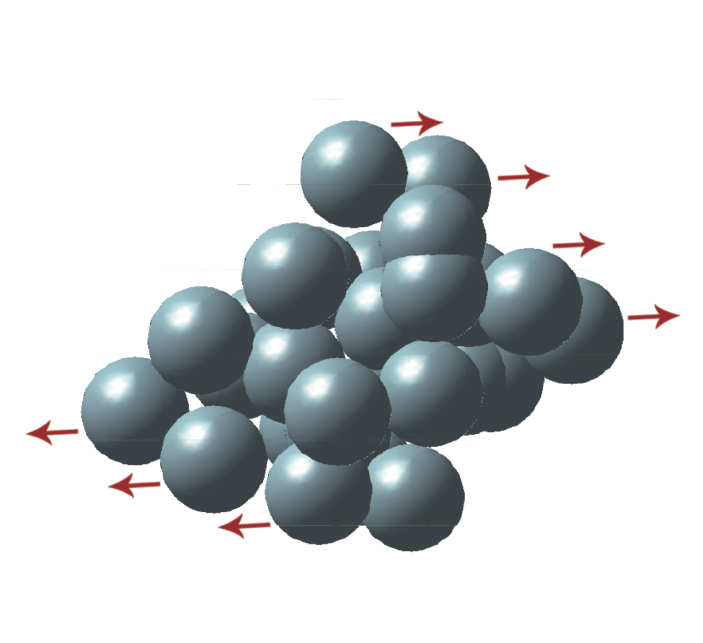} 
 }
\caption{$8$ representative forms of disturbances corresponding to eigenvectors $\phi_i$. The snapshots are taken in CPF, and the reconfiguring effects are exaggerated for better visibility. The small arrows denote the displacement of spheres, the thick arrows denote coherent translation of the cluster, and the curved arrows denote the coherent rotation of the cluster.}
\label{f:dismode}
\end{figure}

In summary, the eigenvalue set $\left \{ \lambda_i \right \}$ actually measures the stability and stability margin of a configuration for a given cluster. In \fig{eigenvalue}, most $\lambda_i$ are located at the stable branches and only one is located at the unstable branch, with a very small real part, so this largely indicates a stable configuration in CPF. Furthermore, the eigenvector $\phi_i$ associated with $\lambda_i$ describes the type of disturbance corresponding to $\lambda_i$. We examined the $162$ eigenvectors by exaggerating the magnitude of the disturbances. \Fig{dismode} illustrates several typical and representative types of disturbances indicated by these eigenvectors. 

It is notable that the disturbances shown in \fig{dismode} are not based on a complete categorization; instead, they are chosen as a summarry of typical reconfiguring patterns. \Eg\ cases (c)--(g), each represent tens of different $\phi_i$, with similar disturbance types. Specifically, (a) indicates speeding up the rotation, corresponding to $6$ pure imaginary $\lambda_i$, \ie\ the critical stable state of a rotating cluster; with similar disturbance types (b) denotes the rigid translation mode corresponding to $6$ zero $\lambda_i$, indicating the clusters under translational disturbance are all stable; (c)--(g) denote complex $\lambda_i$ at stable branches, including simple tension/compression (c, d), and coupled tension and compression (e, f, g), all produced in reflection symmetry; type (h) denotes the shearing mode corresponding to an $\lambda_i$ at the unstable branch, suggesting that the cluster would be much more susceptible under shearing disturbance compared to other types. Also, the magnitude of $\lambda_i$ in stable branches suggests the extent of structural stability of the cluster, \eg\ the compression (d) has $\lambda_i$ with larger magnitude than the tension (c) and accordingly the compressed cluster gains a faster recovery from the disturbance. 

As a related aside, this example is not deliberately chosen. As a random equilibrium cluster, it provides a simplistic model to enable the eigenvector analysis, and interesting results can be presented in an intuitional way. The number of the disturbance forms depends highly on the interior structure of the equilibrium cluster (the force chain), and things will go complex when a size distribution of the spheres is considered. As a usual means to verify the eigen analysis, the numeric results meet the theoretical predictions well, which is the basis for applying this method in the following data processing of \sect{result}.

\section{Numerical Experiments} \label{s:experi}

\subsection{Cluster Generating} \label{s:gener}

The gravitational $N$-body code \code{pkdgrav} was used for numerical experiments in this study. The code has been adapted to treat both soft-sphere and hard-sphere collisions for planetesimal modeling using a high-performance gravity-tree algorithm [10,23]. 

The first group of simulations aims at generating as many different configurations as possible, in order to determine the range of structures of the cluster of the same components. We start with a disperse cloud of $27$ particles as specified in \sect{examp}, which are initially dispersed in a $560\times640\times400$ m cuboid, and allow it to collapse under its self-gravity. The initial positions and velocities of these particles are random, subject to a given total angular momentum $\mathbf{L}$. For each run, the simulation is performed until the aggregate of particles settles down, \ie\ an equilibrated cluster is generated. A total mass $8.24\times10^6$ ton and overall dimension $\sim 300$ m were chosen to be representative of a small asteroid, \eg\ asteroid (99942) Apophis. Three $\mathbf{L}$ values were used to construct the clusters; minimum angular momentum $\mathbf{L}=0$ kg$\cdot$m$^2\cdot$s$^{-1}$ (no spin); medium angular momentum $\mathbf{L}=5.5164\times10^9$ kg$\cdot$m$^2\cdot$s$^{-1}$ (period $9$--$16$ h); and maximum angular momentum $\mathbf{L}=1.1033\times10^{10}$ kg$\cdot$m$^2\cdot$s$^{-1}$ (period $4$--$7$ h), which is a high spin rate for rubble-pile asteroids in practice. $3000$ clusters were generated for each $\mathbf{L}$ following this approach. 

The created clusters with different interior structures serve our reference set, based on which we observe the connections between the structural properties and the eigenvalues' spectrum $\lambda_i$ statistically. By studying these widely distributed cluster samples for three different $\mathbf{L}$ values, we seek a confirmation of the regimes revealed by the eigen analysis and a better understanding of the dependence of the cluster's resistance on the configuration.

\subsection{Impact Simulations} \label{s:collis}

The reshaping processes induced by non-disruptive impacts are the focus of this study. These impacts from micrometeorites are regarded as a trivial low-energy events experienced by most asteroids larger than $100$ m. Since a rubble-pile asteroid can reshape following a regime rearrangement of the constituent blocks, we reproduce these processes with numerical simulations by modeling collisional effects on the sample clusters.  

In the second group of simulations, instantaneous impulses are applied to the clusters generated in \sect{gener} to mimic the impacts from moderate meteorites. Special attention is paid to the bulk reconfiguration of sample clusters, thus we ignore the details of the impact on the surface such as the cratering mechanism and the ejecta formation, although these non-catastrophic impacts are reasonably of high energy. It aims at a principle exploration, and the details of the interaction between the projectile and the asteroid's surface must be handled with care in a subsequent discussion. The impulse is given an arbitrary direction and limited magnitude (to avoid disruption) to mimic a single impact from a micrometeorite. The impacting location is randomly chosen among the surface particles of the cluster. The traveling waves among the media of constituent blocks are numerically approximated by subsequent interactions between the particles, which is basically a principle approximation to the sound propagation because we employed parameters biased away from the rock material. The simulation continues until the cluster equilibrates again, with only two possible outcomes: an irreversible rearrangement that changes the cluster's structure; or reversion to the original configuration (\ie\ a stable outcome; see \sect{examp}). This process is repeated automatically for each simulation until a very robust configuration is reached for the specified impact magnitude. The interval between two impulses needs to be long enough to ensure that the cluster can settle down completely. The magnitude of the impulse is chosen based on the surface seismograms in previous research of (433) Eros's surface modification process [24] that produced a seismic wave in magnitude of $1$ mm at hundreds of meters from the impact site. Thus a magnitude $1.87\times10^7$ kg$\cdot$m$\cdot$s$^{-1}$ is used, which is very large considering the small total mass of the example cluster. It will reproduce a seismic shaking of significant magnitude at positions far from the impacted site.

A simple and efficient indicator is required to trace the reshaping processes of each cluster (tracking all the $9000$ clusters using the coordinates of configuration $\mathbf{x}$ would be a cumbersome task). Simply measuring the semi-axes of the enclosing ellipsoid is not sufficient, because clusters of tens of particles have relatively strong shearing strength that prevents the overall shape coming to fluid equilibrium shapes. Instead, we adopt the cluster's three principal moments of inertia $\left [ I_x, I_y, I_z \right ]$ as indicators to quantify the clusters (in CPF), and we introduce the volume of the inertia ellipsoid $V_\textup{E}$ to indicate the degree of disorder of the particles' arrangement in the cluster, or packing density: 

\begin{equation} 
\label{e:Vdef}
V_\textup{E} = I_x I_y I_z.
\end{equation}

The third group of simulations consists of a grid search on two principal parameters, the rotational speed, \ie\ $\mathbf{L}$, and the impact magnitude. The three clusters in \fig{inerdistri} are representative of the simulation outcomes and cover a wide range of $V_\textup{E}$: A is $3.1512$ $10^{40}$ kg$^3\cdot$m$^6$ (hexagonal packing), B is $3.7161$ $10^{40}$ kg$^3\cdot$m$^6$ (moderately disordered packing) and C is $4.2806$ $10^{40}$ kg$^3\cdot$m$^6$ (very disordered packing). For each cluster, the simulations are conducted at the nodes of a $6\times6$ mesh grid, which is parameterized with rotational speed from $0.0$ to $2.67\times10^{-4}$ rad$\cdot$s$^{-1}$ and impact magnitude from $0.0$ to $1.87 \times 10^7$ kg$\cdot$m$\cdot$s$^{-1}$. For each simulation, the disturbing impacts are repeated again and again until the configuration reaches a final state that no longer responds to the impacts. The final configurations of the clusters are recorded and the changes are quantified by $V_\textup{E}$. 

\section{Results} \label{s:result}

\subsection{Cluster Statistics} \label{s:statis}

The sample clusters provide an approximation to the representative set of all possible configurations. In this section, the clusters are represented in CPF. We scale these clusters by layering their coordinates $\left [ I_x, I_y, I_z \right ]$ on the level surfaces of $V_\textup{E}$, thus a basic law for the distribution is derived. Then we calculate several basic physical properties of these clusters, and by establishing connections between these properties, we show the stable indicator derived from eigen analysis is globally valid for our experiments. 

\subsubsection{Distribution of $\left [ I_x, I_y, I_z \right ]$} \label{s:inerdis}

\begin{figure}[h]
\centering
\scalebox{0.32}
{\includegraphics{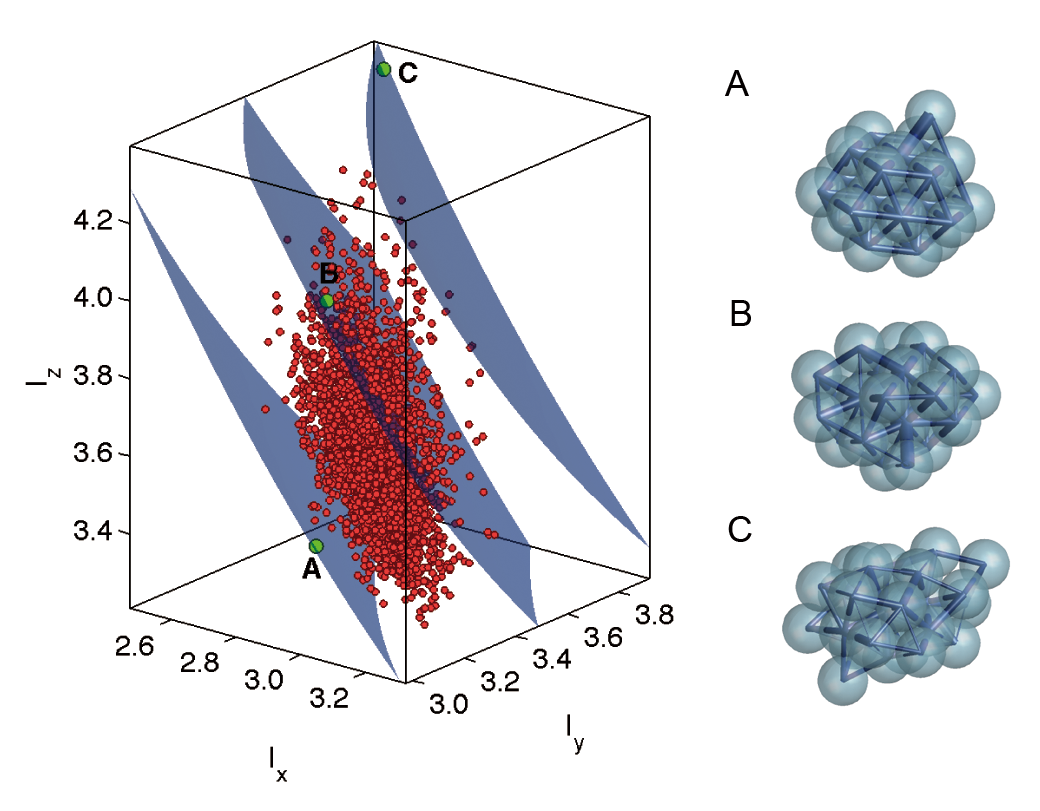}}
\caption{The distribution of $3000$ clusters with $\mathbf{L}=0$ in the inertia frame (unit: $10^{13}$ kg$\cdot$m$^2$). Each of the red solid circles indicates the three coordinates $I_x, I_y, I_z$ of the corresponding cluster. The blue sheets indicate level surfaces of $V_\textup{E}$, and the labeled green markers on them indicate the three sample clusters A, B, C, which are demonstrated with the snapshots on the side. The bubbles indicate the particles in the cluster A, B and C, and the dark frameworks indicate the contacts between the particles, which are shown to illustrate their configurations.}
\label{f:inerdistri}
\end{figure}

\Fig{inerdistri} illustrates the $3000$ clusters of zero total angular momentum (minimum $\mathbf{L}$) in the inertia frame. The diffuse distribution reveals the enormous number of possible configurations even for a cluster of only $27$ particles, which is attributed to the multiple degrees of freedom. It is worth noticing that the ideal equilibrium structures are isolate in the coordinate system, because the static friction is omitted in our model. 

$V_\textup{E}$ serves as a useful indicator for the clusters' overall shapes because it is an integral parameter that correlates with the arrangement of the cluster. In general, clusters in dense packing (low porosity) have smaller $V_\textup{E}$ than those in loose packing (high porosity). To take a detailed look at the structure of the distributed ``cloud" in \fig{inerdistri}, we present three level surfaces of $V_\textup{E}$ that denote small, medium and large values of $V_\textup{E}$. Three clusters A, B and C are indicated on the three respective surfaces, so that the configurations of clusters at different levels can be checked visually: cluster A in the bottom of the cloud represents almost hexagonal packing, which is a dense regular packing with a maximum volume fraction ($\sim0.64$); cluster B is on the middle layer where the cloud is heavily populated and moderately disordered; cluster C is at the sparse area at the top of the cloud; it is highly disordered and represents a prolate overall shape. 

\begin{figure}[h]
\centering
\scalebox{0.21}
{\includegraphics{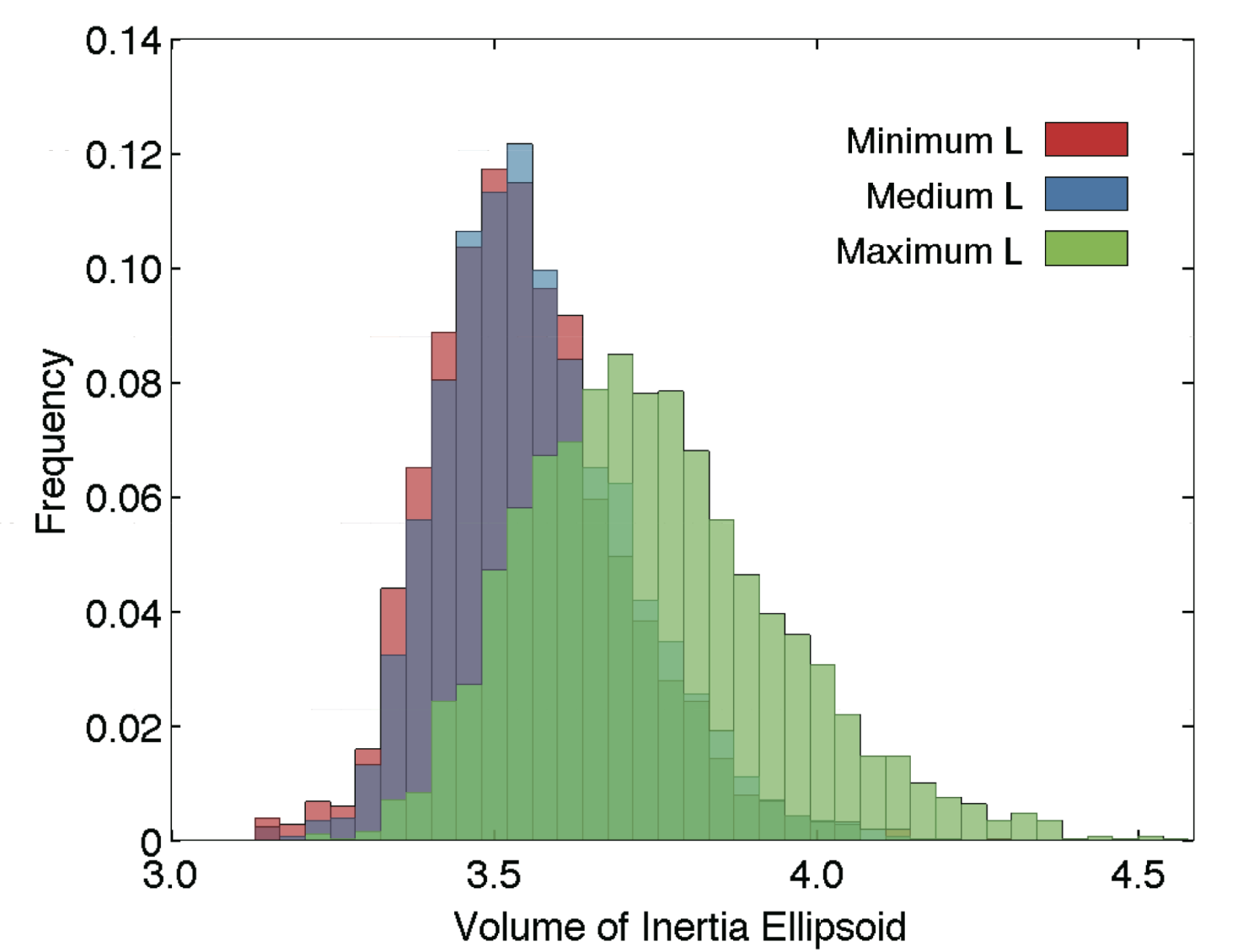}}
\caption{The histograms of $V_\textup{E}$ for three sets of clusters with different total angular momentum clusters (unit: $10^{40}$ kg$^3\cdot$m$^6$). Each set includes $3000$ clusters (see \sect{gener}). The red, blue and green bars indicate the frequency of clusters at minimum, medium and maximum $\mathbf{L}$, respectively.}
\label{f:histogram}
\end{figure}

The distribution in \fig{inerdistri} suggests that natural aggregation of boulders tend to evolve into configurations with moderate disorder, which is a hypothesis of no external disturbances based on probability. Another finding is, the cluster's extent of disorder (or porosity) grows as $V_\textup{E}$ increasing. \Fig{histogram} shows the histograms of all three sets of clusters with different total angular momentum $\mathbf{L}$ (see \sect{gener}). For each set, we calculate the frequency of clusters with different $V_\textup{E}$. The clusters at minimum $\mathbf{L}$ present a bias-normal distribution with relatively low covariance; the medium $\mathbf{L}$ results in a slight positive shift of the histogram than the minimum $\mathbf{L}$; and the maximum $\mathbf{L}$ causes a significant positive shift of the histogram and more decentralized distribution. 

\Fig{histogram} shows the probability distribution of clusters' configurations generated formed randomly dispersed particles, that the most likely outcome is neither dense packing nor loose disordered packing. This result suggests that the most stable structure is difficult to form by once disturbance. After a dispersive impact, the reaccumulated constituents most likely form a new aggregate in sub-stable state. The very robust cluster must be created by repeated moderate disturbances, which can break all unstable links between the particles. Besides, \fig{histogram} also shows the rotation plays an important role in the reshaping processes of the cluster, which will be discussed in \sect{depend}.

\subsubsection{Structural factors $V_E$, $P_G$ and $I_S$} \label{s:correl}

\begin{figure}[h]
\centering
 \subfigure[Gravitational potential $P_\textup{G}$ varying as $V_\textup{E}$.]{
   \label{}
   \includegraphics[width=0.6\textwidth] {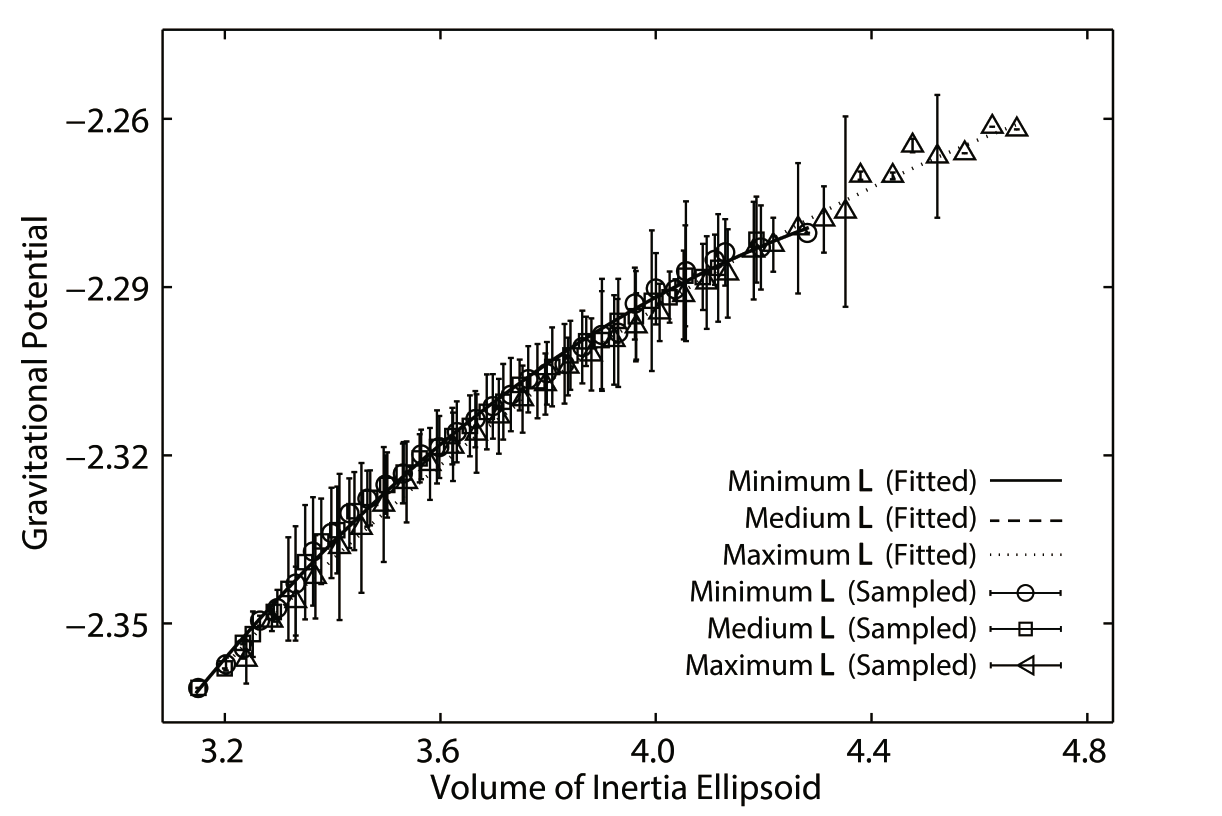}
 }
 \subfigure[Stability index $I_\textup{S}$ varying as $V_\textup{E}$.]{
   \label{}
   \includegraphics[width=0.6\textwidth] {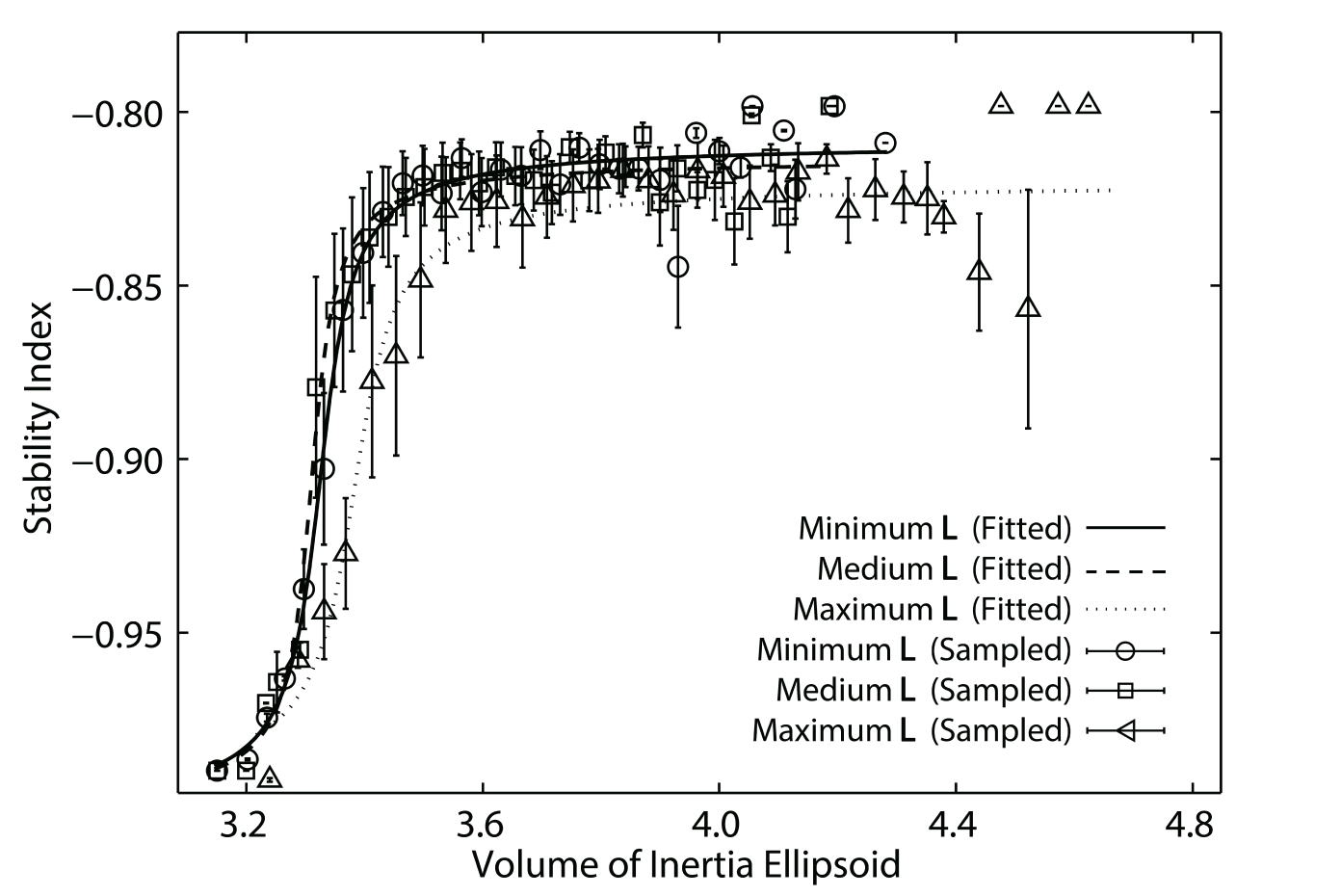} 
 }
\caption{(a) Mutual gravitational potential $P_\textup{G}$ (unit: $10^7$ kg$\cdot$m$^2\cdot$s$^{-2}$) and (b) stability index $I_\textup{S}$ (unit: s$^{-1}$) distribution plots of the clusters for three $\mathbf{L}$ values (see \sect{gener}). The errorbars indicate the $1-\sigma$ deviations of the stability index at each level. The solid, dashed and dotted lines indicate the fitted curves for the three $\mathbf{L}$ values.}
\label{f:potenstab}
\end{figure}

As noted in \sect{inerdis}, the structural stability (or strength) of the cluster is highly related to the arrangement of the particles, which is generally indicated by the volume of inertia ellipsoid $V_\textup{E}$. The total gravitational potential $P_\textup{G}$ partly explains this connection, that the porous clusters (large $V_\textup{E}$) store more energy than clusters with dense packing, which is the reason for the latter usually has better stability under external disturbances. On the other hand, the spectum of eigenvalues $\lambda_i$ also denotes the cluster's structural stability (see \sect{eigen}). We define the stability index of a cluster $I_\textup{S}$ by

\begin{equation} 
\label{e:indstab}
I_\textup{S} = \sum_{i} {\textup{Re} \left ( \lambda_i \right )}, 
\end{equation}

\noindent which reflects the cluster's total capability to resist external disturbances. Thus we get three indicators highly correlated with the structural strength: $V_\textup{E}$, $P_\textup{G}$ and $I_\textup{S}$, which are derived separately based on different theories or simulations. \Fig{potenstab} shows the fitting curves of $P_\textup{G}$ (a) and $I_\textup{S}$ (b) varying as $V_\textup{E}$ based on $9000$ sample clusters of three $\mathbf{L}$ values (see \sect{gener}). As illustrated, $P_\textup{G}$ increases uniformly as $V_\textup{E}$ in nearly linear fashion, suggesting a monotonic connection between the interior degree of disorder and the extra potential energy restored in the cluster. $I_\textup{S}$ shows a steep increase with $V_\textup{E}$ then asymptotes when $V_\textup{E} > 3.5\times10^{40}$ kg$^3\cdot$m$^6$, implying a highly unstable margin where the clusters are disordered in configuration. 

Unlike the $P_\textup{G}$ curves, which are almost the same at all three $\mathbf{L}$ values,  $I_\textup{S}$ shows the influence of the total angular momentum: medium $\mathbf{L}$ makes not much difference, which is consistent with the probability distribution in \fig{histogram}, while high $\mathbf{L}$ leads to a significant shift down in the fitted curve, implying that high spin rate is capable to enhance the resistance of a cluster under external disturbances. The reason is the centrifugal force partly counteracts the self-gravity of the cluster and prevents it from collapsing towards low $V_\textup{E}$.

In summary, we find $V_\textup{E}$, $P_\textup{G}$ and $I_\textup{S}$ are correlated variables, which suggests they are good indicators of the cluster's structural stability, and shows the connection between structural strength and particle arrangement, namely that the strength is reduced as the interior degree of disorder increases. The rotation also plays an important role, with a high proportion of the loose disordered clusters are obtained at maximum $\mathbf{L}$, showing the enhancing effects of the centrifugal force. These features are fitted well with the results of a recent study on the structure of the primary of binary Near-Earth asteroid ($65803$) Didymos [25]. In this work, they found the structured interior provides greater shear strength, which may be enough to explain the primary's fast rotation without the need for cohesion. 

\subsection{Structural Evolution due to Impacts} \label{s:evolut}

\begin{figure}[h]
\centering
 \subfigure[The changes of the clusters' $I_x, I_y, I_z$ due to impacts.]{
   \label{}
   \includegraphics[width=0.48\textwidth] {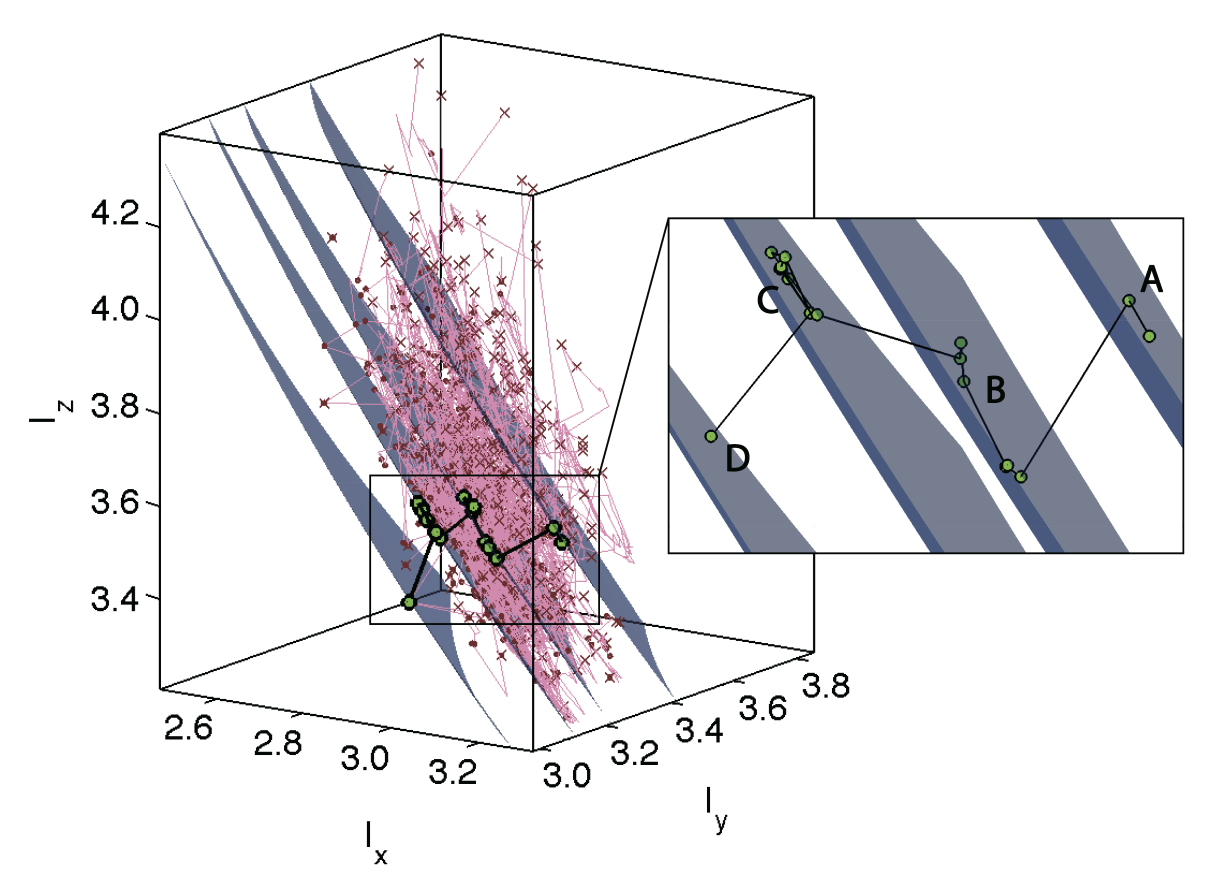}
 }
 \subfigure[The variation of $V_\textup{E}$ with the number of impacts.]{
   \label{}
   \includegraphics[width=0.48\textwidth] {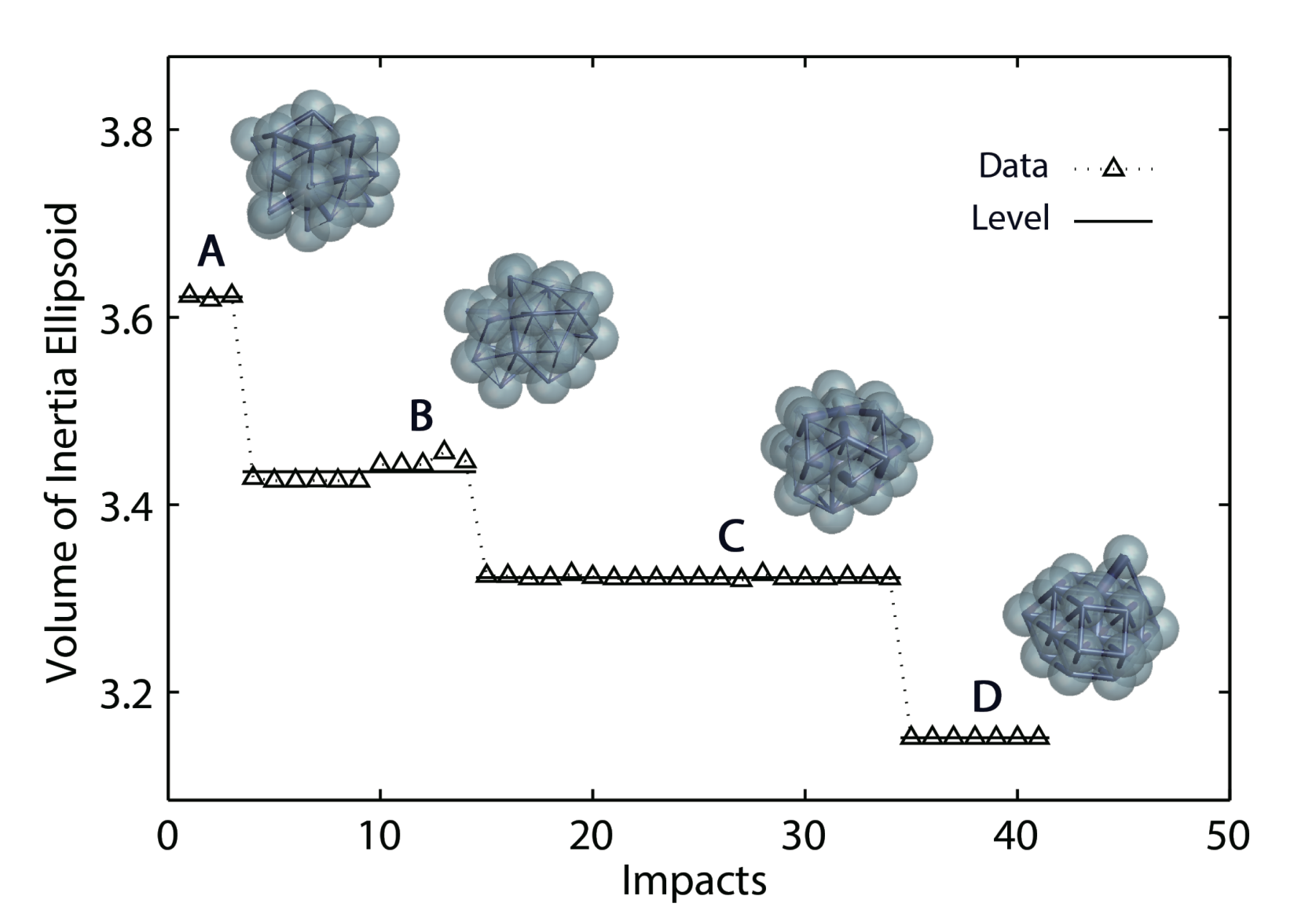} 
 }
\caption{(a) The changes of the clusters' $\left [ I_x, I_y, I_z \right ]$ due to impacts, based on $3000$ clusters with $\mathbf{L}=0$ (see \sect{gener}). Fifty impacts of fixed magnitude are applied to ensure the clusters can rarely be reshaped any further. The crosses indicate the original coordinates of each cluster, and the dots indicate their final coordinates. The light solid lines indicate the paths between the original and final configurations. The dark solid line indicates a representative cases among these paths, on which all the intermediate structures of the cluster are marked with solid green circles. Four blue sheets indicate the level surfaces of $V_\textup{E}$ where the cluster is at the ``wander" stage, and structures on them are marked with A, B, C, D respectively, presented in an enlarged view. (b) The variation of $V_\textup{E}$ with the number of impacts for the highlighted cluster shown in (a). The triangle markers indicate $V_\textup{E}$ values at each impact, and solid lines indicate the levels of the four sets. Four bubble frameworks are shown to represent the structure of the cluster at impact $2$ (set A), $11$ (set B), $30$ (set C) and $40$ (set D).}
\label{f:shapemap}
\end{figure}

\Fig{shapemap}(a) records the changes of $\left [ I_x, I_y, I_z \right ]$ for clusters of $\mathbf{L}=0$ during a series of random collisional processes described in \sect{collis}. All the clusters are confirmed to reach robust equilibria before $50$ impacts, \ie\ the fixed-magnitude impulse does not change their configurations any further. The structural evolution of each cluster is recorded and illustrated by inertia paths (\fig{shapemap}). A common trend that can be observed among the $3000$ samples is all clusters exhibit non-increasing $V_\textup{E}$, and most exhibit a step-wise decrease in $V_\textup{E}$. To take a detailed look at the structural changes, we highlight one of the paths and show it in the enlarged box. Four level surfaces of $V_\textup{E}$ are also included at the step values of $V_\textup{E}$. Since the path is chosen randomly, it represents the typical mode of the structural evolution among all samples. We found a common regime that we call a ``wander-jump" during these processes, that among the impacts, the very efficient ones at reducing $V_\textup{E}$ are quite rare. Between two efficient ones, most impacts just manifest as small structural perturbations at the same level surface of $V_\textup{E}$. The small perturbations cause slight reshaping effects in narrow regions in the inertia frame, and sometimes a cyclic path on the level surface. 

\Fig{shapemap}(b) illustrates the variation of $V_\textup{E}$ with the number of impacts, according to the highlighted path shown in \fig{shapemap}(a). The $V_\textup{E}$ decrease is primarily triggered by three impacts, the $3^{rd}$, $14^{th}$ and $34^{th}$. Four isovalue sets of structures are labeled on the highlighted path respectively as A, B, C, D; each represents several structures in the ``wander" stage. We choose one typical structure out of each set and present them with bubble frameworks in the figure. The structural evolution of this cluster exhibits a clear tendency towards low $V_\textup{E}$, \ie\ low gravitational potential and small stability index (see \sect{correl}). The arrangement of the particles trends to be more and more regular and ends up near dense hexagonal packing in set D. This variation indicates an increasing structural strength of the cluster, and \fig{shapemap}(b) also shows an increasing number of impacts is required to further reduce $V_\textup{E}$ from set A to set D, which is numerical evidence for increasing structural strength.

\begin{figure}[h]
\centering
\scalebox{0.29}
{\includegraphics{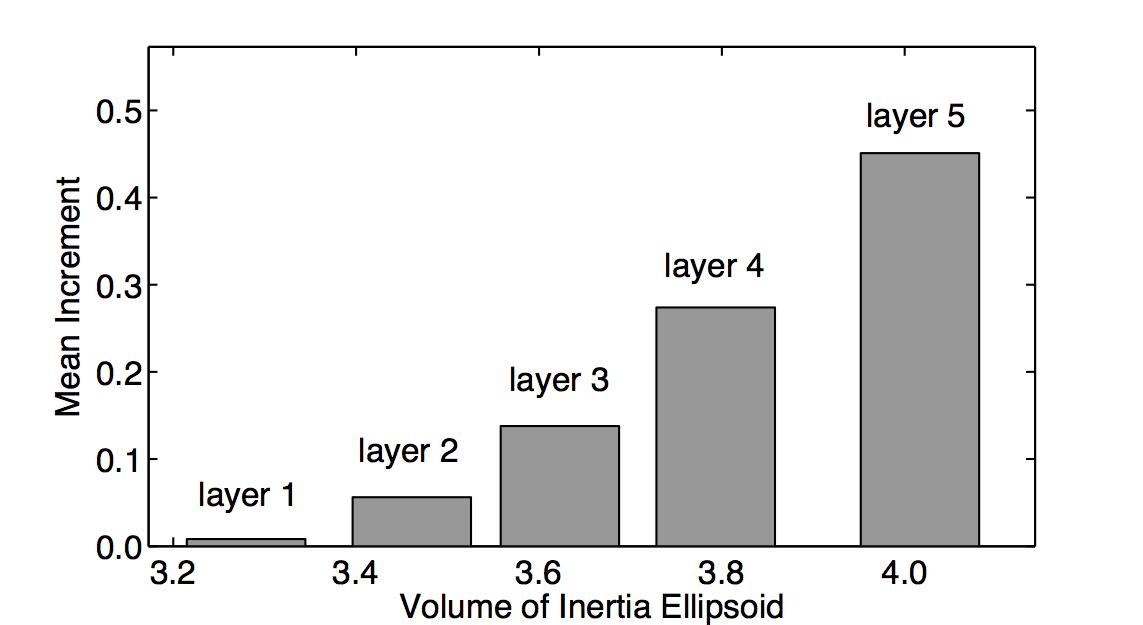}}
\caption{Mean increment of $V_\textup{E}$ in different bins. Five bins are defined on a linear scale of $V_\textup{E}$=$3.1512$, $3.3473$, $3.5434$, $3.7396$, $3.9357$ and $4.1319$ (unit: $10^{40}$ kg$^3\cdot$m$^6$).}
\label{f:increment}
\end{figure}

Returning to rubble-pile asteroids, the results suggest the trivial non-disruptive impacts might play an important role in the rearrangement of the constituent blocks, which may strengthen these rubble piles and help to build a robust structure under impacts of similar magnitude. The dependence of structural changes on $V_\textup{E}$ can be examined by dividing up their evolution int equal bins in $V_\textup{E}$. \Fig{increment} illustrates the clusters divided into five bins in $V_\textup{E}$, showing that larger $V_\textup{E}$ values reduce more quickly in response to impacts. The histogram shows that highly disordered clusters are less capable to resist external disturbances. All the clusters are driven towards low $V_\textup{E}$ by the frequent impacts, and the final mean values for clusters in all five bins drop below $3.5\times10^{40}$ kg$^3\cdot$m$^6$ after the impacts, which is consistent with the criteria predicted by \fig{potenstab}(b) (see \sect{correl}). 

\begin{figure}[h]
\centering
 \subfigure[Minimum $\mathbf{L}$.]{
   \label{}
   \includegraphics[width=0.4\textwidth] {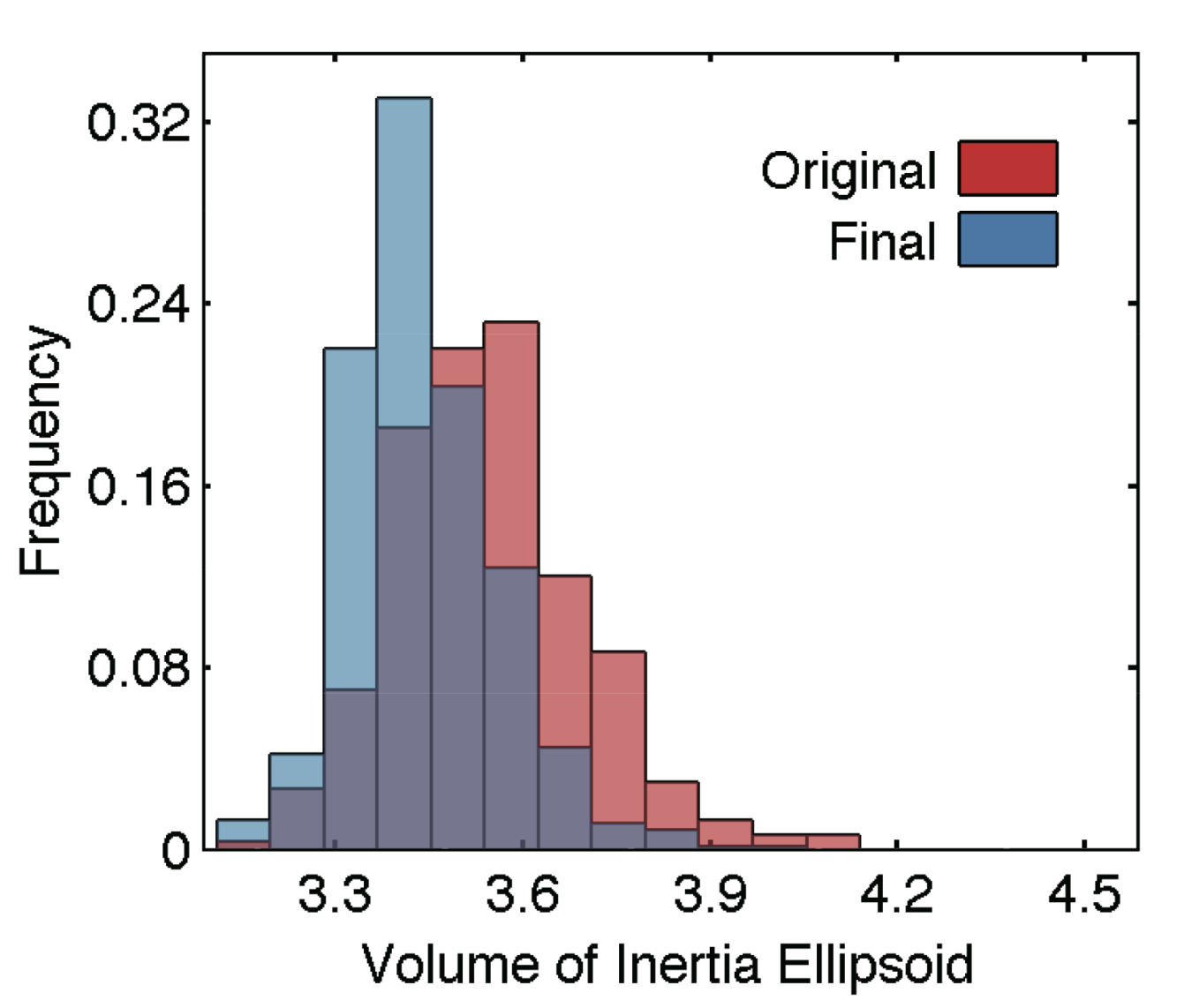}
 }
 \subfigure[Medium $\mathbf{L}$.]{
   \label{}
   \includegraphics[width=0.4\textwidth] {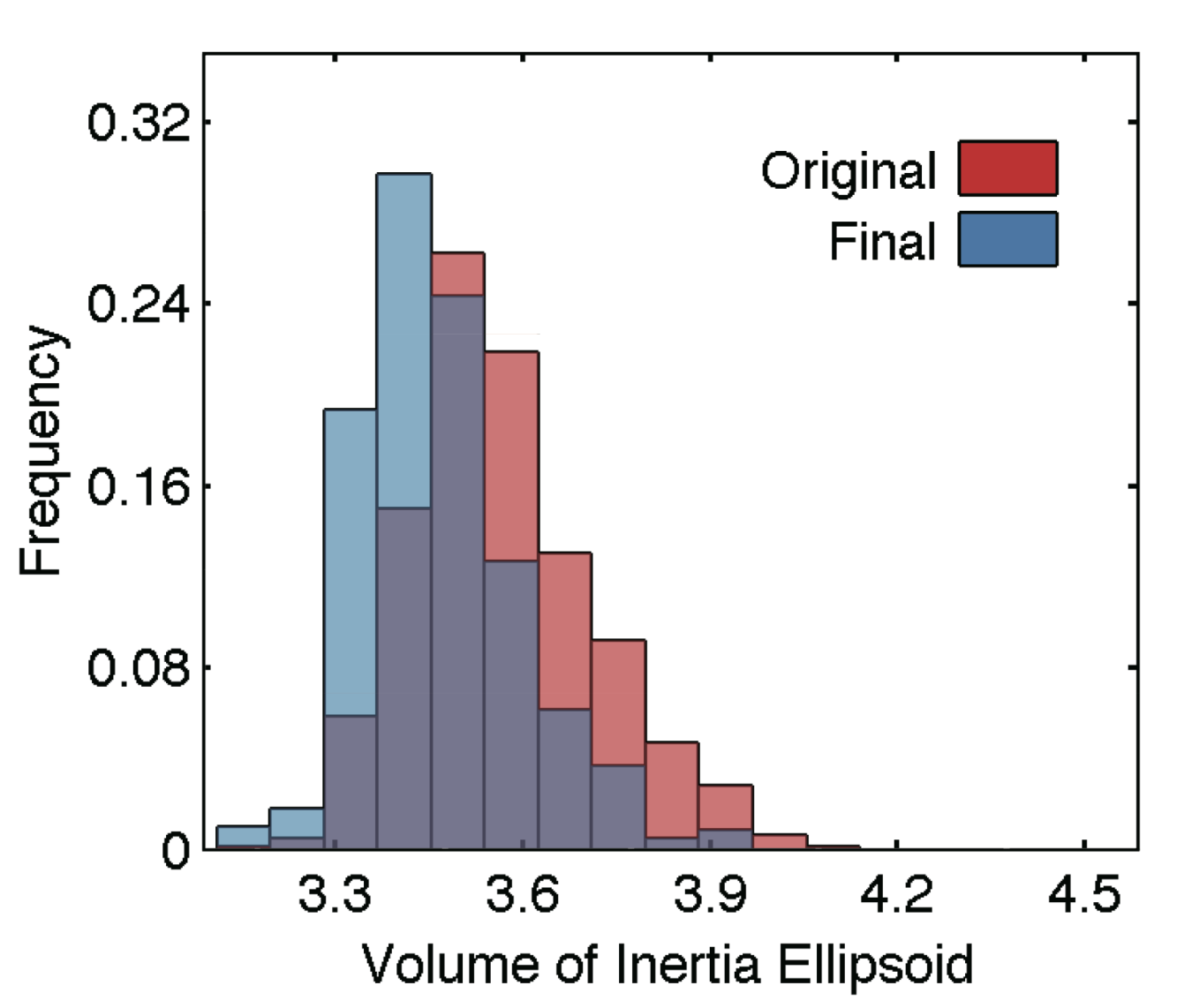} 
 }
 \subfigure[Maximum $\mathbf{L}$.]{
   \label{}
   \includegraphics[width=0.4\textwidth] {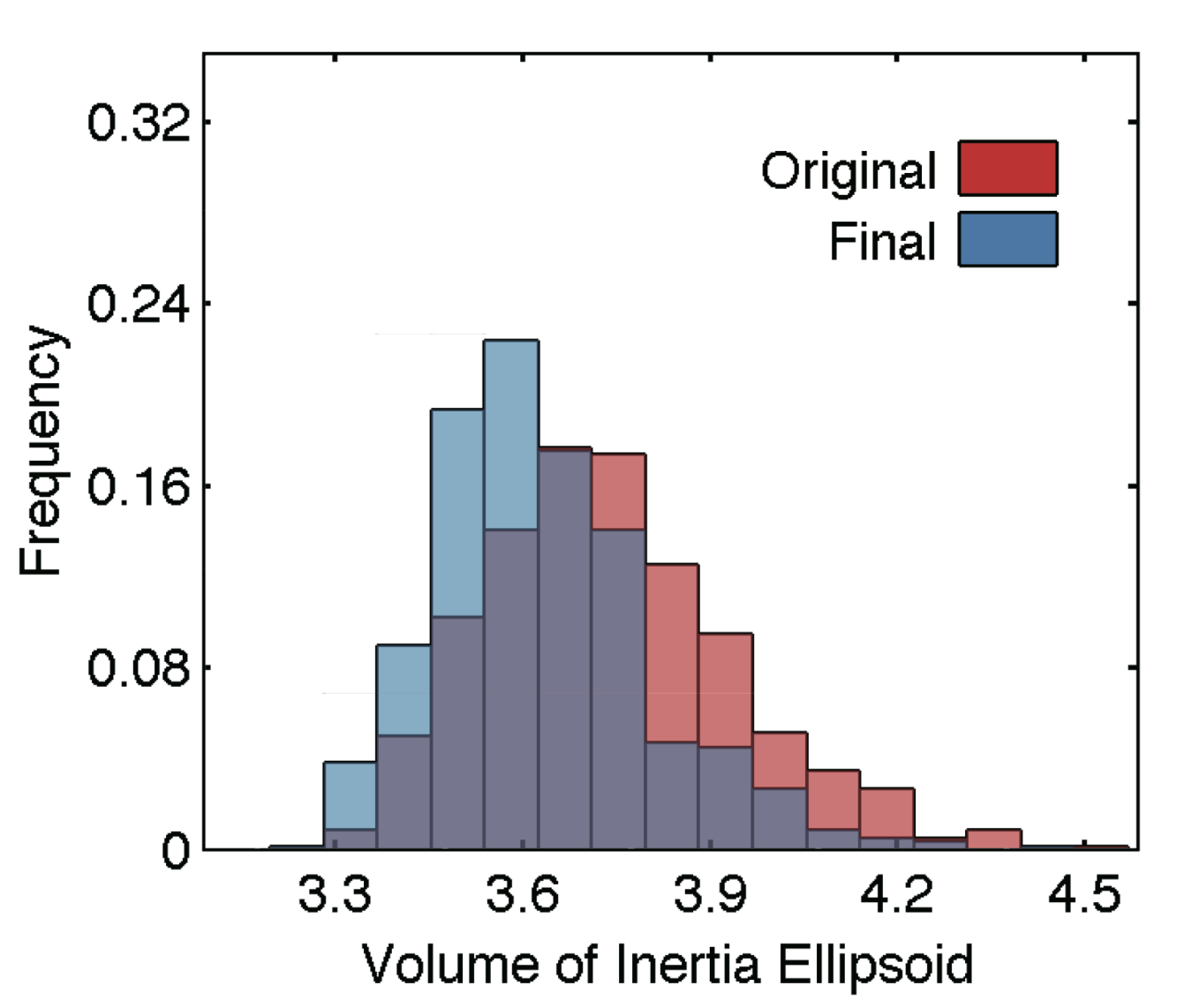} 
 }
\caption{Histograms of $V_\textup{E}$ for clusters at minimum, medium and maximum $\mathbf{L}$ (unit: $10^{40}$ kg$^3\cdot$m$^6$). The red and blue bars indicate clusters before and after the impacts, respectively.}
\label{f:hists}
\end{figure}

\Fig{hists} illustrates the histograms for clusters of three $\mathbf{L}$ values before and after the impacts as a function of $V_\textup{E}$. The final distributions become more concentrated compared to the original ones; while the rotation seems to partly resist the tendency towards dense packing under impacts. The mean value and covariance of these distributions as shown in \tbl{meancov}. 

\begin{table}[h]
  \centering
  \caption{ The mean value and covariance of the distributions in \fig{hists}. }
  \label{t:meancov}
  \begin{tabular}{c|ccc}
    \hline
    $\mathbf{L}$ & State & Mean Value & Covariance\\
    \hline
    Minimum & Original & $3.5459$ & $0.1536$ \\
                    & Final     & $3.4386$ & $0.1184$ \\
    \hline
    Medium   & Original & $3.5664$ & $0.1485$ \\
                    & Final     & $3.4641$ & $0.1261$ \\
    \hline
    Maximum & Original & $3.7370$ & $0.2008$ \\
                     & Final     & $3.6251$ & $0.1714$ \\
    \hline
  \end{tabular}
\end{table}

To interpret what happens to real rubble-pile asteroids, we consider these results in light of current theories regarding the formation of rubble piles. If a rubble pile formed by reaccumulating from fragments of some catastrophic impact, it may suffer continual low-energy impacts during its subsequent evolution that result in shaking and rearrangement of the constituent blocks. The simulations show that porous structures can damp the impact energy very efficiently by reshaping to some dense packing structures with better stability. On the other hand, the total angular momentum plays a significant role during this process, for larger spin rate can prevent the porous clusters from collapsing  down to dense packing, which may be one of the reasons for the fact that several fast-spin asteroids possess relatively low bulk density, such as (15) Eunomia, (45) Eugenia, (87) Sylvia, (762) Pulcova, (121) Hermione, (16) Psyche etc. [26].

\subsection{Impact Parameter Dependence} \label{s:depend}

\begin{figure}[h]
\centering
\scalebox{0.29}
{\includegraphics{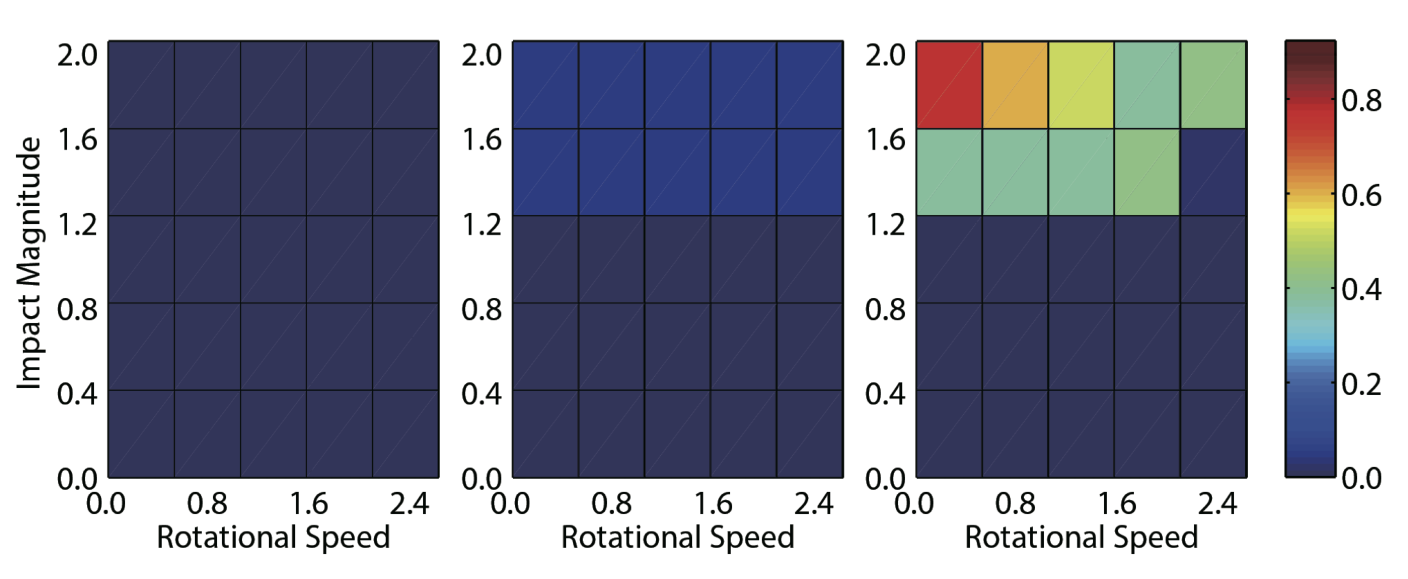}}
\caption{Colormap of the increment of $V_\textup{E}$ for three sample clusters, based on the results from gridded simulations parameterized by rotational speed (unit: $10^{-4}$ rad$\cdot$s$^{-1}$) and impact magnitude (unit: $10^7$ kg$\cdot$m$\cdot$s$^{-1}$). The three clusters demonstrated in \fig{inerdistri} are used: cluster A (left), cluster B (middle) and cluster C (right). A uniform colormap shows the increment of $V_\textup{E}$ in $10^{40}$ kg$^3\cdot$m$^6$.}
\label{f:clrmap}
\end{figure}

\Fig{clrmap} summarizes the results of a grid search on impacts. The diagram of rotational speed and impact magnitude shows the increments of $V_\textup{E}$ at $36$ grid points for the three clusters. It shows these continual non-disruptive impacts do change the clusters' configurations in a common regime. The three cluster configurations A, B, C as identified in \sect{inerdis} are employed. Generally, the dense packing cluster A shows greater robust stability than disordered clusters B and C at all grid nodes, \ie\ the changes of $V_\textup{E}$ for A are much smaller than those for B and C. Rotation proves to be a significant factor in influencing the reshaping effects, as indicated by the fact that the increment of $V_\textup{E}$ decreases as the rotational speed increases, which is quite apparent in \fig{clrmap} for the porous cluster C. It also confirms that more porous interiors may exist among fast-rotating rubble piles compared to the slow-rotating ones, as stated in \sect{evolut}. Also, as the impact magnitude increases, the increment of $V_\textup{E}$ grows for all the three clusters. In particular, we find a magnitude criterion around $1.13\times10^7$ kg$\cdot$m$\cdot$s$^{-1}$ below which all the clusters are almost immune to the impacts. The figure is also consistent with the eigen analysis, for the theory predicts the existence of a stable margin, \ie\ below some magnitude all disturbances around equilibrium can be recovered from eventually.

As an aside, the numerical experiments in this paper are all based on an oversimplified model (soft-sphere clusters) and somewhat arbitrary setup of parameters, which are far from enough to mimic a real astronomical scenario. We present the results as a representation of some qualitative rules, which is a bridge to understanding the common driving mechanisms during the evolution of rubble-pile asteroids. 

\section{Conclusions} \label{s:concl}

In this paper, eigen analysis was introduced to study the structural stability of equilibrated soft-sphere clusters, serving as a representative model for rubble-pile asteroids, so that the strength of a cluster and its responses to external disturbances can be estimated just from its configuration. Further, the results were applied to the analysis of the impact responses of clusters, to explore the dependence of rubble-pile asteroids' reshaping process on the interior structure. Numerical experiments were designed to test our theoretical predictions and to reveal the particular forms of this dependence. 

The eigen analysis shows that the eigenvalues $\lambda_i$ of the equilibrated cluster system indicate its stability to specific types of external disturbances, and corresponding reshaping patterns can be determined using the associated eigenvectors $\phi_i$. Simulations were performed to confirm this finding numerically. 

Randomly generated clusters provided a representative region of possible configurations, with the results showing that the volume of inertia ellipsoid $V_\textup{E}$ is a significant parameter for clusters since it largely measures the degree of disorder of the constituents' arrangement. Statistically, we find the stability index $I_\textup{S}$, total gravitational potential $P_\textup{G}$ and $V_\textup{E}$ are always consistent, and discussed the dependence of the structural stability on these parameters based on analysis of the sample clusters. 

Simulations of the reshaping processes of clusters under non-dispersive impacts reveal a common regime for the structural changes, which we called ``wander-jump", \ie\ the progenitors formed by reaccumulation of initial violent catastrophic impacts may not have very stable interior structures, and a series of collapses triggered by subsequent continual low-energy impacts from small fragments will change the constituents' arrangement gradually towards a dense packing, in a secular geological process. Using the soft-sphere cluster model, we show that when the repeated impacts reach some critical magnitude, the configurations of the samples will evolve consistently towards crystallization after long enough (for equal-sized spheres, the crystallization is hexagonal packing). 

A grid search was performed to identify the parameter dependence of the reshaping effects, showing that the dense packing cluster has the best structural stability among all the samples. The rotational speed proves to play an important role for it exhibits an enhancing effect on loose-packed clusters, \ie\ the centrifugal force partly counteracts the self-gravity and keeps the porous clusters in robust equilibrium, which might explain in part why several fast-spinning asteroids have rather low bulk density. The simulations also suggest a critical magnitude exists, below which all the clusters become immune to external disturbances in any from, consistent with the prediction of the existence of a stable margin from eigen analysis. 

\section*{Acknowledgments}

Y.Y. thanks Prof. H. Baoyin of Tsinghua University for the beneficial discussions. Most of the simulations in this study were run on the YORP computing clusters at the Department of Astronomy, University of Maryland at College Park. 

\section*{References}

\begin{description}

\item{[1]} \aIII {Richardson, D. C., Leinhardt, Z. M., Melosh, H. J., Bottke, J. W. F., Asphaug, E.} {Gravitational aggregates: evidence and evolution} 501--515
\item{[2]} \paper {Bottke, W. F. Jr., Richardson, D. C., Michel, P., Love, S. G.} 1999 {1620 Geographos and 433 Eros: Shaped by planetary tides} {Astron. J.} 117 1921--1928
\item{[3]} \paper {Solem, J. C., Hills, J. G.} 1996 {Shaping of Earth-crossing asteroids by tidal forces} {Astron. J.} 111 1382--1387
\item{[4]} \chap {Chapman C. R.} 1978 {Asteroid collisions, craters, regolith, and lifetimes} {Morrison, D. \etal} {Asteroids: An Exploration Assessment} {NASA Conf. Publ. 2053.} 145--160
\item{[5]} \paper {Weissman, P. R.} 1986 {Are cometary nuclei primordial rubble piles} Nature 320 242--244
\item{[6]} \paper {Benz, W., Asphaug, E.} 1999 {Catastrophic disruptions revisited} Icarus 142 5--20
\item{[7]} \paper {Michel, P., Benz, W., Tanga, P., Richardson, D. C.} 2001 {Collisions and gravitational reaccumulation: Forming asteroid families and satellites} Science 294 1696--1700
\item{[8]} \paper {Asphaug, E., Benz, W.} 1994 {Density of Comet Shoemaker-Levy 9 deduced by modelling breakup of the parent ``rubble pile"} Nature 370 120--124
\item{[9]} \chap {Chodas, P. W., Yeomans, D. K.} 1996 {The orbital motion and impact circumstances of Comet Shoemaker-Levy 9} {Noll, K. S. \etal} {The Collision of Comet Shoemaker-Levy 9 and Jupiter} {Proc. IAU Colloq. 156, Cambridge Univ., Cambridge.} 1--30
\item{[10]} \paper {Richardson, D. C., Bottke, W. F. Jr., Love, S. G.} 1998 {Tidal distortion and disruption of Earth crossing asteroids} Icarus 134 47--76
\item{[11]} \paper {Ballouz, R.-L., Richardson, D.C., Michel, P., Schwartz, S.R., Yu, Y.} 2014 {Numerical simulations of collisional disruption of rotating gravitational aggregates: Dependence on material properties} {Planetary and Space Science} 107 29--35
\item{[12]} \paper {Bottke, W. F. Jr., Richardson, D. C., Love, S. G.} 1997 {Can tidal disruption of asteroids make crater chains on the Earth and Moon} Icarus 126 470--474
\item{[13]} \paper {Farinella, P., Paolicchi, P., Tedesco, E. F., Zappal\`a, V.} 1981 {Triaxial equilibrium ellipsoids among the asteroids} Icarus 46 114--123
\item{[14]} \paper {Kryszczynska, A., La Spina, A., Paolicchi, P., Harris, A. W., Breiter, S., Pravec, P.} 2007 {New findings on asteroid spin-vector distributions} Icarus 192 223--237
\item{[15]} \paper {Richardson, D.C., Elankumaran, P., Sanderson, R.E.} 2005 {Numerical experiments with rubble piles: Equilibrium shapes and spins} Icarus 173 349--361
\item{[16]} \paper {Asphaug, E.} 2010 {Similar-sized collisions and the diversity of planets} {Chemie der Erde - Geochemistry} 70 199--219
\item{[17]} \paper {Tanga, P., Comito, C., Paolicchi, P., Hestroffer, D., Cellino, A., Dell'Oro, A., Richardson, D. C., Walsh, K. J., Delb\`o, M.} 2009 {Rubble-pile reshaping reproduces overall asteroid shapes} {The Astrophysical Journal} 706 197--202
\item{[18]} \paper {S\'anchez, P., Scheeres, D. J.} 2011 {Simulating asteroid rubble piles with a self-gravitating soft-sphere distinct element method model} {The Astrophysical Journal} 727 120--133
\item{[19]} \paper {Schwartz, S. R., Richardson, D. C., Michel, P.} 2012 {An implementation of the soft-sphere discrete element method in a high-performance parallel gravity tree-code} {Granular Matter} 14 363--380
\item{[20]} \book {Perko, L.} 1991 {Differential equations and dynamical systems} {Springer-Verlag, New York}
\item{[21]} \book {Wiggins, S.} 2003 {Introduction to applied nonlinear dynamical systems and chaos} {Springer, New York}
\item{[22]} \book {Shilnikov, L. P., Shilnikov, A. L., Turaev, D. V., Chua, L. O.} 1998 {Methods of qualitative theory in nonlinear dynamics} {World Scientific Publishing Co Inc, Hackensack}
\item{[23]} \chap {Stadel, J., Wadsley, J., Richardson, D.C.} 2002 {High performance computational astrophysics with pkdgrav/gasoline} {Dimopoulos, N. J. \etal} {High Performance Computing Systems and Applications} {Kluwer Academic Publishers, Boston} 501--523
\item{[24]} \paper {Richardson, J. K., Melosh, H. J., Greenberg, R.} 2004 {Impact-induced seismic activity on Asteroid 433 Eros: a surface modification process} Science 306 1524--1529
\item{[25]} \submitted {Zhang, Y., Richardson, D. C., Barnouin, O. S., \etal} 2016 {Geodynamic stability of the proposed AIDA mission target 65803 Didymos: I. Discrete cohesionless granular physics model} {Icarus}
\item{[26]} \aIII {Britt, D. T., Yeomans, D., Housen, K., Consolmagno, G.} {Asteroid density, porosity, and structure} 485--500

\end{description}

\end{document}